\documentclass[lettersize,journal,twoside]{IEEEtran}
\usepackage{amsmath,amssymb,amsfonts}
\usepackage[linesnumbered, ruled]{algorithm2e}
\usepackage{array}
\usepackage[caption=false,font=normalsize,labelfont=sf,textfont=sf]{subfig}
\usepackage{textcomp}
\usepackage{stfloats}
\usepackage{url}
\usepackage{makecell}
\usepackage{verbatim}
\usepackage{graphicx}
\usepackage{cite}
\usepackage{subfig}
\usepackage{graphicx,color}
\usepackage{multirow}
\usepackage{textcomp}
\usepackage{booktabs}
\usepackage{balance}
\usepackage{color}
\usepackage[thmmarks,amsmath]{ntheorem}
\theoremseparator{:}

\theoremheaderfont{\it}
\theorembodyfont{\normalfont}
\theoremsymbol{\ensuremath{\Box}}

\theoremheaderfont{\it}
\theorembodyfont{\normalfont}
\theoremsymbol{}
\newtheorem*{remark}{Remark}
\begin{document}

\title{Deep Reinforcement Learning-Empowered Wireless Sensor Networking for 6G Closed-Loop Controls}

\author{Chengleyang Lei, Wei Feng,~\IEEEmembership{Senior Member,~IEEE}, Yunfei Chen,~\IEEEmembership{Fellow,~IEEE}, Yongxu Zhu,~\IEEEmembership{Senior Member,~IEEE}, Ning Ge,~\IEEEmembership{Member,~IEEE}, and
	Shi~Jin,~\IEEEmembership{Fellow,~IEEE} 
\thanks{C. Lei, W. Feng, and N. Ge are with the Department of Electronic Engineering, State Key Laboratory of Space Network and Communications, Tsinghua University, Beijing 100084, China (email: lcly21@mails.tsinghua.edu.cn, fengwei@tsinghua.edu.cn, gening@tsinghua.edu.cn).}
	\thanks{Y. Chen is with the Department of Engineering, University of Durham, DH1 3LE Durham, U.K. (e-mail: yunfei.chen@durham.ac.uk).}
\thanks{Y. Zhu and S. Jin are with the National Mobile Communications Research Laboratory, Southeast University, Nanjing 210096, China (email: yongxu.zhu@seu.edu.cn, jinshi@seu.edu.cn).}
}




\maketitle

\begin{abstract}   
Robots are increasingly deployed in remote or hazardous areas for mission-critical control tasks. Due to their limited individual capabilities, they have to rely on other field sensors to obtain the state information of targets, and also a dedicated edge information hub (EIH) to enable information exchange, sensing data analysis and control command generation. Such configuration follows a sensing-communication-computing-control ($\textbf{SC}^3$) closed loop. To optimize the whole closed-loop performance, this paper minimizes the linear quadratic regulator (LQR) control cost by designing the sensor-to-EIH bandwidth allocation. Specifically, we first model the distortion noise caused by limited communication data rate based on the mutual information theory. Next, under the control policy based on the Kalman filter and LQR controller, we formulate the control process as a partially observable Markov decision process (POMDP), and develop a deep reinforcement learning (DRL)-based sensor-to-EIH bandwidth allocation scheme. The proximal policy optimization (PPO) algorithm is utilized to train the DRL agent. Simulation results are provided to show the superiority of the proposed DRL-based scheme.
\end{abstract}


\section{Introduction}
\label{Introduction}
\IEEEpubidadjcol
\IEEEPARstart{T}{hanks} to the advances in robotics and artificial intelligence, autonomous robots are increasingly deployed in remote areas to perform mission-critical tasks, such as emergency rescue~\cite{robot1}, scientific exploration, and oil extraction~\cite{robot2}. These robots usually rely on sensors to obtain global environmental information and communication networks to enable information exchange~\cite{network}. However, the terrestrial infrastructures are usually unavailable to serve robots in remote areas due to the harsh geographical conditions. In such scenarios, autonomous aerial vehicles (AAVs, also known as UAVs) can be equipped with devices integrating communication and computing modules that serve as edge information hubs (EIHs) to provide edge intelligence for the robots~\cite{Feng2024}. In addition, a digital twin can be deployed on the EIH, as a virtual representation of the physical control system~\cite{DT1,DT2}. During the control process, the sensors collect environmental information and transmit it to the EIH. Subsequently, the EIH processes this data to analyze the current state of the target, update the target digital twin model, and generate corresponding control commands. These control commands are then transmitted to the robots to guide their actions. The above process is performed periodically, forming a sensing-communication-computing-control ($\textbf{SC}^3$) closed loop~\cite{jsac}. The overall performance and reliability of the robotic tasks depend critically on the efficiency and integrity of the $\textbf{SC}^3$ closed loop. 

Due to the inherent limitation on the payload capacity of UAVs, the communication resources available within the network are usually constrained, posing significant challenges for supporting $\textbf{SC}^3$ closed loops. Moreover, the sensing precision and range of a single sensor are usually limited. This necessitates the deployment of multiple collaborative sensors within an $\textbf{SC}^3$ closed loop to obtain global environmental information~\cite{sensor_network}. The resource allocation among these sensors directly influences the quality of sensing data and accuracy of the digital twin, which in turn fundamentally impacts the overall performance of the $\textbf{SC}^3$ closed loop. Current resource allocation methods for sensors mainly focus on either sensing or communication metrics. However, in an $\textbf{SC}^3$ closed loop, the sensing, communication, computing, and control components cooperate closely to accomplish a common task. The coupling among different components necessitates a comprehensive perspective on the whole $\textbf{SC}^3$ closed loop. Focusing on individual sensing or communication metrics may lead to an insufficient utilization of the limited resources and result in an unsatisfactory performance of the $\textbf{SC}^3$ closed loop. 

Motivated by the above issues, in this paper, we investigate the bandwidth allocation among multiple collaborating sensors within an EIH-assisted $\textbf{SC}^3$ closed loop from a closed-loop perspective. The overall closed-loop control performance of the $\textbf{SC}^3$ closed loop is taken as the optimization objective, and the whole closed-loop control process is comprehensively considered in the bandwidth allocation.

\subsection{Related Works}
In the field of sensing and control, researchers have investigated sensor data fusion algorithms for data collected from multiple sensors. Specifically, the Kalman filter and its variants, which estimate unknown variables based on a series of sensor observations, have been widely used in sensor fusion~\cite{sensor_fusion1}. In \cite{kalman1}, three distributed Kalman filtering algorithms were proposed for sensor networks, and the communication complexity and packet-loss issues were discussed. The authors in \cite{kalman2} proposed a Kalman filter-based approach for human motion tracking which fuses data from a three-axis accelerometer and gyroscope. Recently, with the development of deep learning, many deep-learning-based sensor fusion algorithms have been proposed to fuse multimodal data. Reference \cite{multi-modal1} proposed a deep learning architecture, the Deep Coupling Autoencoder (DCAE), for fault diagnosis in mechanical systems using multimodal data including vibration and acoustic signals collected from sensors. In \cite{multi-modal2}, the authors introduced a multi-modal fusion transformer framework to improve autonomous driving performance in the presence of a high density of dynamic agents and complex scenarios by integrating image and LiDAR data using attention mechanisms. All of these works are valuable for investigating cooperative sensing. However, these studies mainly focus on the design of sensor data fusion algorithms and do not consider the communication components in the $\textbf{SC}^3$ closed loop.

Recently, integrated sensing and communications (ISAC) is recognized as a promising technology. In \cite{isac1}, the authors investigated the partially-connected hybrid beamforming design for multi-user ISAC systems. The digital and analog beamformers were jointly optimized to minimize the Cramér-Rao Bound (CRB) for direction of arrival estimation while ensuring signal-to-interference-plus-noise ratio (SINR) constraints for communication. Reference \cite{isac2} jointly designed the transmit and receive beamformers at the transceiver, as well as the transmit precoder at the uplink user and the receive combiner at the downlink user, to maximize communication rates and radar beampattern power while suppressing residual self interference. In \cite{isac3}, the authors investigated the near-field effect in ISAC. A channel model of the near-field ISAC framework was proposed, based on which the authors optimized the ISAC signal to maximize the sensing performance. However, these works mainly focus on the beamformer design, without considering communication resource allocation.

Several studies have investigated the communication resource allocation among multiple sensors~\cite{sensor1, sensor2, sensor3,sensor5}. The authors in \cite{sensor1} proposed an architecture design of the green wireless sensor networks by exploiting the collaborative energy and information transfer protocol. The subcarrier grouping, subcarrier pairing, and power allocation were jointly optimized to maximize the transmission rate. Reference \cite{sensor2} investigated a wireless powered sensor network, where a power station provides power to sensor nodes to enable their data transmission. The energy beamforming and time allocation were optimized to maximize the sum throughput of the sensor network. In \cite{sensor3}, a joint bandwidth and energy allocation problem was investigated to maximize the total transmit rate from the sensors to the UAV. The authors proposed an optimal algorithm based on dynamic programming. These studies provide valuable insights into resource allocation among multiple sensors. {The authors in \cite{sensor5} investigated partial spectrum sharing in the underwater acoustic sensor network. The minimum data rate was maximized through joint power allocation and spectrum assignment. A hybrid model-based and data-based resource allocation algorithm was proposed to solve the problem.} However, most of them focus on the aim of improving the communication performance, such as the transmission rate. They have not considered the coupling among communication, sensing, and control components within an $\textbf{SC}^3$ closed loop, which prevents the most effective use of information.

There have also been some studies exploring the resource allocation among sensors, considering sensing or control metrics. For example, the authors in \cite{sensor4} investigated the bandwidth allocation problem based on the quantized Kalman filter, aiming at improving the sensing accuracy. However, it modeled the communication as a simple pipe, without considering the channel fading. In \cite{jsac2}, the authors allocated the bandwidth among the sensor-to-UAV and UAV-to-actuator links to minimize the sum linear quadratic regulator (LQR) control cost of multiple $\textbf{SC}^3$ closed loops. However, \cite{jsac2} investigated multiple parallel $\textbf{SC}^3$ closed loops, without considering the collaboration among multiple sensors. {In \cite{sensor6}, the authors investigated the joint sensing and controlling scheduling in an industrial cyber-physical system containing multiple sensors and actuators. It jointly considers the sensing, communication, and control processes, providing useful insights for the study of control-oriented multi-sensor resource scheduling. However, it mainly focuses on transmission scheduling, while bandwidth allocation and the distortion of sensing data under finite-data-rate constraints are not considered.}

Recently, deep reinforcement learning (DRL) algorithms have been used to solve complex resource allocation problems. Tan \textit{et al.} proposed a DRL-based joint secure offloading and resource allocation scheme to improve the secrecy performance and resource efficiency in a vehicular edge computing (VEC) network~\cite{DRL2}. The transmit power, the frequency spectrum selection and the computation resource allocation were jointly optimized based on the double deep Q-learning algorithm to minimize the system processing delay. Reference \cite{DRL3} investigated a joint offloading decision, collaboration decision, computing resource allocation and communication resource allocation problem in the multi-user collaborative mobile edge computing network, with the aim of minimizing the total energy consumption under the delay constraint. A two-level alternation method framework based on the heuristic algorithm and deep Q-learning algorithm was proposed to solve the formulated problem. The above works provide inspiration for the utilization of DRL algorithms for solving the complex resource allocation problems in the $\textbf{SC}^3$ closed loop.

\subsection{Main Contributions}
Motivated by the above observations, this paper investigates an $\textbf{SC}^3$ closed loop assisted by a UAV-mounted EIH. We optimize the bandwidth allocation among collaborative sensors so as to improve the overall performance of the closed loop, considering the specific control process. Towards this end, we propose a DRL-based bandwidth allocation framework to minimize the LQR control cost. The main contributions are summarized as follows.
\begin{itemize}
\item {Different from the existing studies on the bandwidth allocation that mainly optimize sensing accuracy or communication throughput, this paper focuses on the closed-loop control performance. Specifically, the whole control process is considered and the LQR cost is adopted as the optimization objective.}
\item We investigate a robot-control system, which contains multiple sensors, a satellite-UAV network, and a robot. The robot is controlled through an $\textbf{SC}^3$ closed loop. The sensing, communication, computing, and control components within the $\textbf{SC}^3$ closed loop are jointly considered.
\item {We establish a cross-layer relation between bandwidth allocation and closed-loop control performance. Specifically, the limited communication data rate is mapped to an equivalent distortion noise variance in the received sensing data, thereby  further affecting the state estimation and the control input generation.} We model the distortion noise as Gaussian and obtain its variance based on mutual information theory. Based on the control policy using the Kalman filter and LQR controller, we propose a DRL-based framework to design bandwidth allocation policy utilizing the proximal policy optimization (PPO) method.
\item We provide simulation results to show that the proposed DRL-based scheme can achieve lower LQR control cost than traditional communication-oriented methods.
\end{itemize}

\subsection{Organization and Notation}
The rest of this paper is organized as follows. Section \ref{sec_system} introduces the system model, formulating the bandwidth allocation design problem. In Section \ref{sec_algorithm}, we propose a DRL-based bandwidth allocation framework. Simulation results are provided in Section \ref{sec_simulation} with further discussions. Finally, Section \ref{sec_conclusion} concludes this paper.

\section{System Model and Problem Formulation}
\label{sec_system}

\begin{figure} [t]
	\centering
	\includegraphics[width=0.9\linewidth]{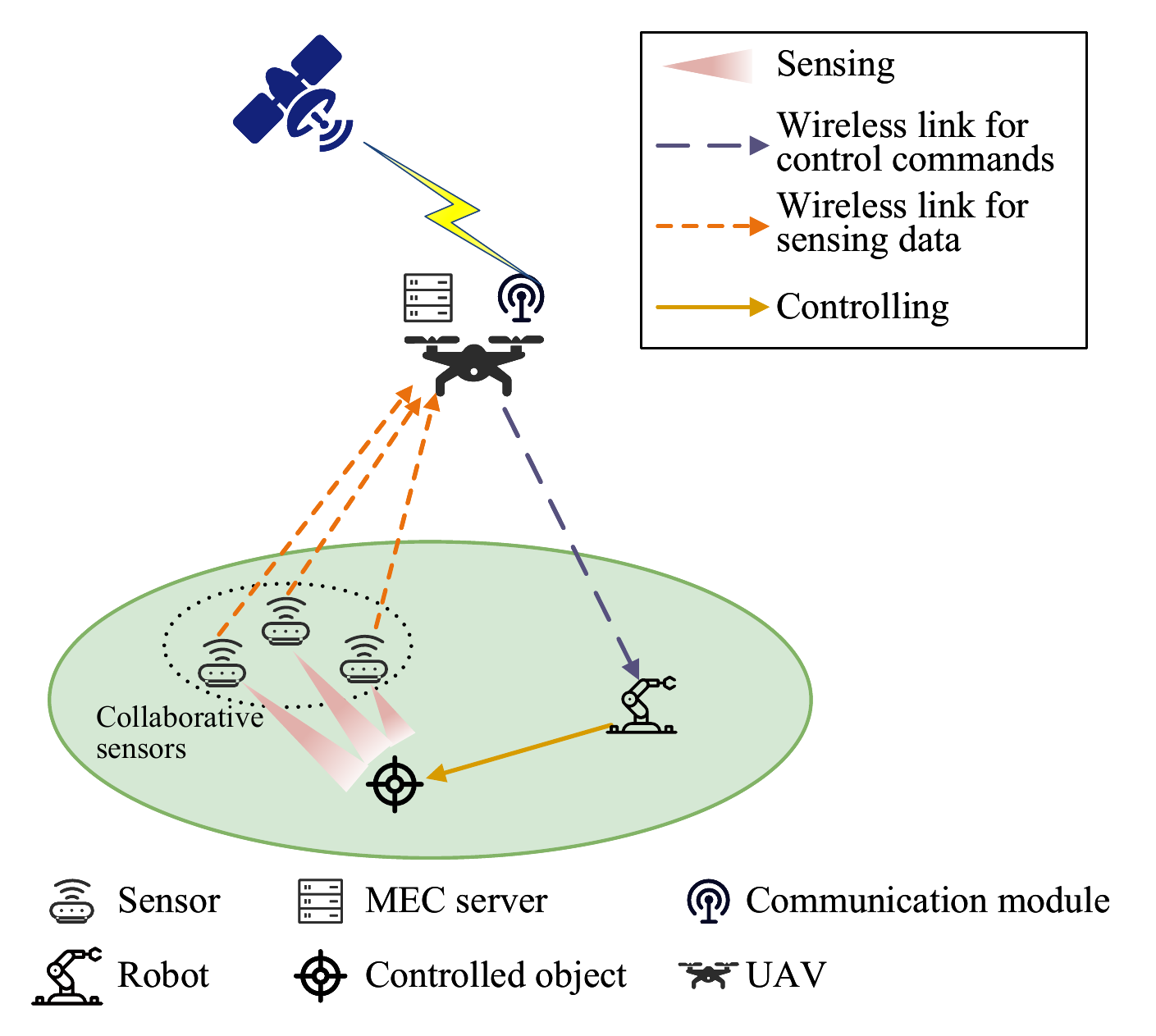}
	\caption{Illustration of the closed-loop control system where multiple sensors cooperate to support the control process, assisted by a satellite-UAV-enabled EIH, which integrates not only the communication module but also the mobile edge computing (MEC) server.}
	\label{fig:system}
\end{figure} 
As shown in Fig. \ref{fig:system}, in this paper, we investigate an $\textbf{SC}^3$ closed loop which contains $K$ sensors, a UAV-mounted EIH\footnote{{The EIH can be considered as an integrated on-site information hub for the $\textbf{SC}^3$ closed loop. It not only provides edge computing capability similar to existing edge computing systems, but also coordinates sensing data collection, communication resource allocation, digital-twin-based state estimation, and control command generation. By regarding the EIH as a unified entity integrating sensing, communication, and computing information collection and analysis, the coupling among the functions within the $\textbf{SC}^3$ closed loop can be better considered\cite{jsac}.}}, and a robot. A satellite is utilized to support the backhaul of the on-site information, as well as the UAV telemetry and tracking. The satellite does not directly participate in the closed-loop control process studied in this paper. The UAV is equipped with an EIH integrating communication and computing modules to support the task, whose position is fixed and known during the task. A digital twin is integrated in the EIH to maintain a virtual representation of the physical control system. Specifically, by analyzing the observed data transmitted from the sensors, the digital twin updates the system state estimate in real time. Based on the updated state estimate, the EIH generates control commands accordingly, which are then transmitted to the robot to guide its actions. This forms an $\textbf{SC}^3$ closed loop. In the following subsections, we introduce detailed models of the sensing, communication, computing, and control components of the $\textbf{SC}^3$ closed loop. For ease of reference, the main notations are listed in Table~\ref{tab1}.

\begin{table}[!t]
	\centering
	\caption{Summary of notations.}
	\label{tab1}
	\begin{tabular}{cp{5.3cm}}\toprule
		\textbf{Notation} & \textbf{Description}\\\midrule
		$\mathbf{B}_t$ & Bandwidth allocation action of the DRL agent in cycle $t$ \\\hline
		$B_{k,t}$ & Bandwidth allocated to sensor $k$ in cycle $t$ \\\hline
		$B_{\text{max}}$ & Bandwidth constraint \\\hline
		$\mathbf{c}_k$ & Observation vector of sensor $k$ \\\hline
		$D_{k,t}$ & Data throughput from sensor $k$ to the EIH at cycle $t$	\\\hline
		$g_{k}$ & Large-scale channel fading from sensor $k$ to the EIH \\\hline
		$h_{k,t}$ & Channel gain from sensor $k$ to the EIH in cycle $t$ \\\hline
		$K$ & Number of sensors	\\\hline
		$\mathbf{K}_{t}$ & Kalman filter gain at cycle $t$	\\\hline
		$l$ & Infinite-horizon LQR cost of the control system\\\hline
		$n$ & Dimension of the system state of the control system	\\\hline
		${n}_{k,t}$ & Distortion of the received sensing result of sensor $k$ in cycle $t$	\\\hline
		${N}_{0}$ & Communication noise power spectral density	\\\hline
		$\mathbf{o}_t$ & Observation of the DRL agent in cycle $t$ \\\hline
		$p_{k,t}$ & Transmit power from sensor $k$to the EIH at cycle $t$	\\\hline
		$\mathbf{P}_{t}$ & Error covariance matrix of the posterior estimate for the control system state in cycle $t$	\\\hline
		$\mathbf{P}_{t|t-1}$ & Error covariance matrix of the prior estimate for the control system state in cycle $t$	\\\hline
		$\text{PL}_{k}$ & Path loss from sensor $k$ to the EIH	\\\hline
		$\mathbf{Q}$,$\mathbf{R}$ & Weight matrices of the LQR cost\\\hline
		$r_t$ & Reward of the DRL agent in cycle $t$ \\\hline
		$T$ 	& Transmission time of sensor data per cycle	\\\hline
		$\mathbf{u}_{t}$ 	& Control input of control system in cycle $t$	\\\hline
		$\mathbf{v}_{t}$ 	& Noise of control system in cycle $t$	\\\hline
		$\mathbf{V}$ 	& Covariance matrix of control system noise	\\\hline
		$w_{k,t}$ 	& Sensing noise for sensor $k$ in cycle $t$	\\\hline
		$\mathbf{x}_{t}$ 	& State of control system in cycle $t$	\\\hline
		$\hat{\mathbf{x}}_{t}$ & Posterior estimate for the control system state based on Kalman filter in cycle $t$	\\\hline
		$\hat{\mathbf{x}}_{t|t-1}$ & Prior estimate for the control system state in cycle $t$	\\\hline
		$y_{k,t}$ 	& Sensing result of sensor $k$ in cycle $t$	\\\hline
		$\widetilde{y}_{k,t}$ 	& Received sensing result at the EIH from sensor $k$ in cycle $t$	\\\hline
		$\mathbf{\Phi}$ & State matrix of the control system \\\hline
		$\mathbf{\Gamma}$ & Input matrix of control system \\\hline
		$\mathbf{\sigma}_{\text{w},k}^2$	&	Sensing noise variance for sensor $k$	\\\hline
		$\rho$	&	Information utilization ratio
		\\\bottomrule
	\end{tabular}
\end{table}

\subsection{Control and Sensing Model}
First, we present the models of the control system and the sensing process. In this paper, without loss of generality, a typical linear system is modeled for control. The discrete-time system equation in cycle $t$ is given by~\cite{control_system}
\begin{equation}\label{system}
	\mathbf{x}_{t+1} = \mathbf{\Phi}\mathbf{x}_{t}+\mathbf{\Gamma}\mathbf{u}_{t}+\mathbf{v}_{t},
\end{equation}
where $\mathbf{x}_{t}\in \mathbb{R}^{n}$ and $\mathbf{u}_{t}\in \mathbb{R}^{m}$ represent the system state and the control input, $n$ and $m$ denote their dimensions, $\mathbf{v}_{t}\in \mathbb{R}^{n}$ denotes the Gaussian control noise with mean zero and covariance $\mathbf{V}$, and $\mathbf{\Phi}\in \mathbb{R}^{n\times n}$ and $\mathbf{\Gamma}\in \mathbb{R}^{n\times m}$ denote the state transition matrix and control input matrix, respectively. We assume that the initial state $\mathbf{x}_{1}$ follows a Gaussian distribution with mean zero and covariance $\mathbf{P}_{1}$.

The sensing process at each sensor is linear. We assume that the observation equation of each sensor is a scalar value, and the sensing process of sensor $k$ is given as
\begin{equation}\label{sensing}
	{y}_{k,t} = \mathbf{c}_k^\text{T} \mathbf{x}_{t} + {w}_{k,t},
\end{equation}
where ${y}_{k,t}$ denotes the output of sensor $k$, $\mathbf{c}_k\in \mathbb{R}^{n}$ is the sensor vector, and ${w}_{k,t}$ denotes the Gaussian sensing noise with mean zero and covariance $\mathbf{\sigma}_{\text{w},k}^2$. 

After observing the state of the control target, the sensors transmit their acquired data to the EIH through wireless channels. Due to the limited communication data rate of the sensor-to-EIH links, the received data suffers from distortion, leading to discrepancies between the received data $\widetilde{y}_{k,t}$ and the original sensing measurements ${y}_{k,t}$. According to \cite{distortion_noise}, when minimizing the accumulated directed information from the sensor to the controller subject to a control cost constraint, the optimal information-reduction mechanism admits a linear–Gaussian realization, i.e., the signal available at the receiver can be represented as a linear transformation of the original signal corrupted by additive white Gaussian noise (AWGN). Inspired by the linear–Gaussian structure in \cite{distortion_noise}, in this paper, we model the rate-limited distortion in the received signal as an equivalent additive Gaussian noise\footnote{While practical communication channels may introduce non-Gaussian distortion, it has been shown that the quantization noises can be well approximated as AWGN in the high-dimensional or small-quantization-step cases~\cite{distortion_noise1, distortion_noise2}.}, referred to as the distortion noise throughout the paper. The received signal of sensor $k$ at the EIH can be modeled as
\begin{equation}\label{distortion}
	\widetilde{y}_{k,t} = 	{y}_{k,t} + {n}_{k,t},
\end{equation}
where ${n}_{k,t}$ is the zero-mean Gaussian distortion noise, whose variance depends on the communication data rate.

Next, we derive the relation between the distortion noise variance and the communication data rate. Assuming that the sensing result ${y}_{k,t}$ is Gaussian with variance $\sigma^2_{{y}_{k,t}}$ and is independent of the distortion noise, the mutual information between the source ${y}_{k,t}$ and the received signal $\widetilde{y}_{k,t}$ can be calculated as
\begin{subequations}\label{mutual_information}
	\begin{align}
	I\left({y}_{k,t}; \widetilde{y}_{k,t}\right) & = h(\widetilde{y}_{k,t})-h(\widetilde{y}_{k,t}\mid y_{k,t})\\& = h(y_{k,t}+n_{k,t})-h(n_{k,t})\\
	&= \frac{1}{2} \log_2 \left( 1 + \frac{\sigma^2_{{y}_{k,t}}}{\sigma^2_{{n}_{k,t}}} \right),
	\end{align}
\end{subequations}
where $\sigma^2_{{n}_{k,t}}$ denotes the variance of ${n}_{k,t}$. We introduce an information utilization ratio $\rho\in(0,1]$ to capture that only a fraction of the raw sensing content is task-relevant to the control objective~\cite{jsac2}. Then the amount of task-relevant information that can be utilized at the EIH is upper bounded by $\rho D_{k,t}$, where $D_{k,t}$ denotes the transmitted data throughput from sensor $k$ to the EIH in cycle $t$. Assuming that the task-relevant information is fully utilized, we have $I\left({y}_{k,t}; \widetilde{y}_{k,t}\right)=\rho D_{k,t}$. The distortion noise variance can be calculated as
\begin{equation}
	\sigma^2_{{n}_{k,t}} = \frac{\sigma^2_{{y}_{k,t}}}{2^{2\rho D_{k,t}}-1}.
\end{equation}
The above equation indicates that the quality of the received data is determined by the communication data rate, which is influenced by the bandwidth allocation among the sensors. Based on the above relation, in this paper, we aim to find the optimal bandwidth allocation policy to improve the overall control performance.
 
\begin{remark}
	The above derivation is based on the assumption that the signal ${y}_{k,t}$ follows a Gaussian distribution. Note that we model all the noises ($\mathbf{v}_{t}$, ${w}_{k,t}$, and ${n}_{k,t}$) as Gaussian random variables, which is a common assumption in control theory. As $\mathbf{x}_{t}$ and ${y}_{k,t}$ are linear combinations of the deterministic parameters and Gaussian random variables, they also follow Gaussian distributions.
\end{remark}

\subsection{Uplink Transmission Model}
The sensors transmit their sensing results to the EIH simultaneously over orthogonal frequency channels, so we assume that there is no interference between the sensor-to-EIH links. We consider a realistic air-to-ground channel model for the sensor-to-EIH links that incorporates both line-of-sight (LoS) and non-line-of-sight (nLoS) elements~\cite{channel}. The channel gain can be expressed as
\begin{align}
h_{k,t}=\sqrt{g_k}s_{k,t},
\end{align}
where $g_k$ denotes the large-scale channel fading, and $s_{k,t}$ denotes the small-scale channel fading. The small-scale channel fading is assumed to be Rayleigh distributed. According to \cite{channel}, the path loss of the channel from sensor $k$ to the EIH can be calculated as
\begin{equation} \label{path_loss}
	\text{PL}_{k}=\frac{A^\text{PL}}{1+ae^{-b(\theta_{k}-a)}}+B^\text{PL}_{k},
\end{equation}
where 
\begin{align}
	&A^\text{PL}=\eta_{\text{LOS}}-\eta_{\text{NLOS}},\\
	&B^\text{PL}_{k}=20\text{log}_{10}(d_{k})+20\text{log}_{10}(\frac{4\pi f}{c})+\eta_{\text{NLOS}}.
\end{align}
Here, $\eta_{\text{LOS}}$, $\eta_{\text{NLOS}}$, $a$ and $b$ are constant parameters related to the propagation environments, $f$ represents the carrier frequency, $c$ denotes the speed of light, $d_{k}$ and $\theta_{k}$ represent the distance and elevation angle between the EIH and sensor $k$ respectively. Then the large-scale channel fading can be expressed as
\begin{equation} \label{large_scale}
	g_{k}=10^{-\text{PL}_{k}/10}.
\end{equation}

Denoting the bandwidth of the channel from sensor $k$ to the EIH in cycle $t$ as $B_{k,t}$, we have
\begin{equation}
	\sum_{k = 1}^{K} B_{k,t} \leq B_{\text{max}},
\end{equation}
where $B_{\text{max}}$ denotes the maximum bandwidth constraint.

Due to the tight requirements of delay in robot control, short packets and finite blocklength should be applied. In such a case, the maximum achievable rate in the finite blocklength regime is approximated by~\cite{short_packet}
\begin{align}
	\begin{split}
		R_{k,t}  \left( B_{k,t} \right)= &B_{k,t} \log_2 \left( 1 + \frac{|h_{k,t}|^2 p_{k,t}}{B_{k,t} N_0}\right) \\ - &\sqrt{\frac{B_{k,t}V_{k,t}}{T}}Q^{-1}\left(\epsilon \right) + \frac{\log_2 \left(B_{k,t} T \right) }{2 T},
	\end{split}
\end{align}
where $T$ denotes the transmission time, $p_{k,t}$ denotes the transmit power, $N_0$ denotes the noise power spectral density, $Q^{-1}\left( \cdot \right) $ denotes the inverse of the Q-function, which is defined as $Q\left( \cdot \right) \triangleq \frac{1}{\sqrt{2\pi}} \int_x^\infty e^{-\frac{t^2}{2}} \, dt $, $\epsilon$ denotes the packet error probability, and $V_{k,t}$ denotes the channel dispersion, which can be expressed as
\begin{equation}
	V_{k,t} = \left( 1 - \frac{1}{(1+\frac{|h_{k,t}|^2 p_{k,t}}{B_{k,t} N_0})^2} \right) \left( \log_2 e \right)^2.
\end{equation}

In this paper, we focus on ultra-reliable operation and set $\epsilon$ to a very small value. Consequently, the sensor-to-EIH links can be modeled as reliable bit-pipes with effective information rate $R_{k,t}(B_{k,t})$. Random packet-drop events are not explicitly simulated in the state evolution. Therefore, the data throughput of sensor $k$ can be calculated as $D_{k,t} = TR_{k,t}\left(B_{k,t} \right)$.

\subsection{Computing and Downlink Transmission Model}
After receiving the sensing results, the EIH analyzes the sensing data, updates the digital twin, and generates control action $\mathbf{u}_t$. The computing process can be expressed using the following function
\begin{align}\label{computing}
	\mathbf{u}_t = f_t \left(\widetilde{Y}^t,  U^{t-1} \right),
\end{align}
where 
\begin{align}
	&\widetilde{Y}^t \triangleq \left[ \widetilde{\mathbf{y}}_1, \widetilde{\mathbf{y}}_2,\cdots, \widetilde{\mathbf{y}}_t \right],\\
	&U^{t-1} =\left[\mathbf{u}_1, \mathbf{u}_2, \cdot, \mathbf{u}_{t-1}  \right], 
\end{align}
with $\widetilde{\mathbf{y}}_t \triangleq \left[\widetilde{y}_{1,t},  \widetilde{y}_{2,t}, \cdots \widetilde{y}_{K,t}\right]^{\text{T}}$.
The equation in \eqref{computing} indicates that the generation of control input $\mathbf{u}_t$ depends not only on the current sensing results $\mathbf{y}_t$, but also on the entire history of the sensing results $\widetilde{Y}^t$ and the past control inputs $U^{t-1}$.

After the computing process, the EIH transmits the control commands to the robot on the ground. Finally, the robot executes the received control commands, finishing the current control cycle. The control command packets from the EIH to the robot are typically much shorter than the raw sensor data that contains redundant observations. Moreover, the downlink transmit power from the EIH is usually much larger than the uplink power at individual sensors. Therefore, in this paper, we assume that the control commands are received by the robot without distortion, and focus on the uplink bandwidth allocation, as the main bottleneck. Incorporating the effects of downlink imperfections into a joint uplink and downlink resource allocation is an interesting direction for future work.

\subsection{Problem Formulation}
The primary objective of this work is to optimize the overall performance of the $\textbf{SC}^3$ closed loop. We utilize the infinite-horizon LQR cost to measure the control performance, which is formulated as~\cite{LQR}
\begin{equation}\label{LQR}
	l \triangleq \lim\limits_{N\rightarrow \infty}\mathbb{E} \left[ \frac{1}{N}\sum_{t = 1}^{N} \left(\mathbf{x}_{t}^\text{T}\mathbf{Q}\mathbf{x}_{t} +\mathbf{u}_{t}^\text{T}\mathbf{R}\mathbf{u}_{t}\right) \right],
\end{equation}
where  $\mathbf{Q}$ and $\mathbf{R}$ are weight matrices. The term $\mathbf{x}_{t}^\text{T}\mathbf{Q}\mathbf{x}_{t}$ measures the deviation of the system from zero state, and the term $\mathbf{u}_{t}^\text{T}\mathbf{R}\mathbf{u}_{t}$ evaluates the control energy consumption. The LQR cost comprehensively measures the state convergence and energy consumption, which is a common index in control theory. A smaller LQR cost indicates a better control performance.

The bandwidth allocation among sensors will influence the distortion noise variances, thereby impacting the state estimation process of the digital twin. In this paper, we design the bandwidth allocation $\mathbf{B}_t \triangleq \left[B_{1,t}, B_{2,t}, \cdots, B_{K,t} \right] $ to minimize the LQR cost of the control system. The optimization problem can be formulated as
\begin{subequations}\label{P1}
	\begin{align}
		\min_{\left\{\mathbf{B}_t \right\}} \ &\lim\limits_{N\rightarrow \infty}\mathbb{E} \left[ \frac{1}{N}\sum_{t = 1}^{N} \left(\mathbf{x}_{t}^\text{T}\mathbf{Q}\mathbf{x}_{t} +\mathbf{u}_{t}^\text{T}\mathbf{R}\mathbf{u}_{t}\right) \right] \label{P1a} \\ 
		\mbox{\textit{s.t.}}\  & \sum_{k = 1}^{K} B_{k,t} \leq B_{\text{max}},\quad \forall t, \label{P1b} 
		\\&B_{k,t}\geq 0, \quad \forall k,t,\label{P1c}
		\\&	\sigma^2_{{n}_{k,t}} =  \frac{\sigma^2_{{y}_{k,t}}}{2^{2\rho T R_{k,t}\left(B_{k,t} \right) }-1} \quad \forall k,t,\label{P1h}
	\end{align}
\end{subequations}
where the control system state evolves as \eqref{system}-\eqref{distortion}, \eqref{computing}. Note that the bandwidth allocation affects the LQR cost through the distortion noise variance in \eqref{distortion}, which further impacts state estimation and control decisions. The above problem is difficult to solve directly due to the expectation over the stochastic noises in the objective function. Moreover, the control policy $f_t$ will influence the LQR cost. In the next section, we first introduce the widely-used separation-based control policy combining a Kalman filter for state estimation and an LQR controller. We then solve the bandwidth allocation problem based on the DRL method.

\section{DRL-Based Bandwidth Allocation for Collaborative Sensing}
\label{sec_algorithm}
\subsection{Control policy based on Kalman filter and LQR controller}
First, we rewrite \eqref{sensing} and \eqref{distortion} as
\begin{equation}\label{re_sensing}
	\widetilde{\mathbf{y}}_{t} = \mathbf{C} \mathbf{x}_{t} + \widetilde{\mathbf{w}}_{t},
\end{equation}
where $\mathbf{C} = \left[\mathbf{c}_1, \mathbf{c}_2, \cdots, \mathbf{c}_K \right]^{\text{T}}  $, and
\begin{align}
	\widetilde{\mathbf{w}}_{t} = \mathbf{w}_t + \mathbf{n}_t ={
	\begin{bmatrix}
		w_{1,t} \\
		\vdots \\
		w_{K,t}
	\end{bmatrix}}
	+
	{\begin{bmatrix}
		n_{1,t} \\
		\vdots \\
		n_{K,t}
	\end{bmatrix}
}.
\end{align}
As $ \mathbf{w}_t$ and $\mathbf{n}_t$ are independent, the covariance of $\widetilde{\mathbf{w}}_{t}$ can be calculated as
\begin{equation}
 	\widetilde{\mathbf{W}}_t = \text{diag}\left\lbrace \sigma^2_{\text{w},k}+ \frac{\sigma^2_{{y}_{k,t}}}{2^{2\rho D_{k,t}}-1} \right\rbrace.
\end{equation}

It can be seen that the revised sensing equation \eqref{re_sensing} has the same form as the classical sensing equation in control theory. As both the system noise $\mathbf{v}_{t}$ and sensing noise $\mathbf{w}_{t}$ are Gaussian variables, the classical linear quadratic Gaussian control (LQG) can be applied~\cite{LQG}. Specifically, we can utilize the Kalman filter to obtain the optimal estimate of the system state, and then obtain the control inputs based on the estimate. Based on the Kalman filter, the optimal estimation of $\mathbf{x}_{t}$ can be calculated as~\cite{Kalman}
\begin{equation}\label{estimation}
	\hat{\mathbf{x}}_{t} = \hat{\mathbf{x}}_{t|t-1} + \mathbf{K}_{t} \left( \widetilde{\mathbf{y}}_{t} -  \mathbf{C}\hat{\mathbf{x}}_{t|t-1} \right),
\end{equation}
where $\hat{\mathbf{x}}_{t}$ is the optimal posterior estimate of $\mathbf{x}_t$ based on the sensing results $\widetilde{\mathbf{y}}_{t}$, $\hat{\mathbf{x}}_{t|t-1}$ denotes the prior estimate before incorporating $\widetilde{\mathbf{y}}_{t}$, $\mathbf{K}_{t}$ denotes the Kalman filter gain, and $\widetilde{\mathbf{y}}_{t} -  \mathbf{C}\hat{\mathbf{x}}_{t|t-1}$ is the innovation or residual, representing the discrepancy between the measurement and prediction. In addition, the posterior estimate error covariance matrix of $\hat{\mathbf{x}}_{t}$, denoted as $\mathbf{P}_{t}$, is calculated as~\cite{Kalman}
\begin{equation}
	\mathbf{P}_{t} =  \left(\mathbf{I} - \mathbf{K}_{t}\mathbf{C} \right) \mathbf{P}_{t|t-1},
\end{equation}
where $\mathbf{P}_{t|t-1}$ is the prior error covariance matrix of $\hat{\mathbf{x}}_{t|t-1}$.

The Kalman filter gain $\mathbf{K}_{t}$ can be obtained by~\cite{Kalman}
\begin{equation}
	\mathbf{K}_{t} = \mathbf{P}_{t|t-1} \mathbf{C}^{\text{T}}  \left(\mathbf{C} \mathbf{P}_{t|t-1}\mathbf{C}^{\text{T}} + \widetilde{\mathbf{W}}_t \right) ^{-1}.
\end{equation}

The prior estimate $\hat{\mathbf{x}}_{t+1|t}$ and its error covariance matrix $\mathbf{P}_{t+1|t}$ in the next control cycle can be calculated recursively, as~\cite{Kalman}
\begin{align}
&\hat{\mathbf{x}}_{t+1|t} = \mathbf{\Phi}\hat{\mathbf{x}}_{t}+\mathbf{\Gamma}\mathbf{u}_{t},\label{update1}\\
&\mathbf{P}_{t+1|t} =\mathbf{\Phi} \mathbf{P}_{t}\mathbf{\Phi}^{\text{T}} + \mathbf{V}.\label{update2}
\end{align}

Based on the Kalman posterior estimate $\hat{\mathbf{x}}_{t}$, the optimal control input can be calculated according to the optimal LQR controller as
\begin{equation}
	\mathbf{u}_{t} = -\mathbf{G}\hat{\mathbf{x}}_{t},
\end{equation}
where $\mathbf{G}$ is the optimal LQR controller gain, which can be calculated as~\cite{LQG}
\begin{equation}
	\mathbf{G} = (\mathbf{R} + \mathbf{\Gamma}^{\text{T}} \mathbf{S} \mathbf{\Gamma})^{-1} \mathbf{\Gamma}^{\text{T}} \mathbf{S} \mathbf{\Phi},
\end{equation}
where $\mathbf{S}$ is the solution to the following Riccati equation
\begin{equation}\label{Riccati}
\mathbf{S} = \mathbf{\Phi}^{\text{T}} \mathbf{S} \mathbf{\Phi} - (\mathbf{\Phi}^{\text{T}} \mathbf{S} \mathbf{\Gamma})(\mathbf{R} + \mathbf{\Gamma}^{\text{T}} \mathbf{S} \mathbf{\Gamma})^{-1} (\mathbf{\Gamma}^{\text{T}} \mathbf{S} \mathbf{\Phi}) + \mathbf{Q}.
\end{equation}

\subsection{Critical Factors of POMDP}
Based on the Kalman filter and LQR controller framework, we can obtain the control input $\mathbf{u}_{t}$ of each control cycle. However, it is still difficult to design an optimal bandwidth allocation policy due to the expectation operator within the objective function of \eqref{P1}. To address this challenge, we propose a data-driven approach based on the DRL method. Within the DRL framework, we can learn an effective bandwidth allocation policy directly from the interaction with the environment. The whole $\textbf{SC}^3$ closed loop can be naturally modeled as a POMDP according to the system dynamics from \eqref{system}, the observation equation \eqref{sensing}, and the estimation and control processes defined in  \eqref{estimation}-\eqref{Riccati}. Next, we introduce the critical factors of the POMDP, including state space, observation space, action space, and reward function. 
\subsubsection{State Space}
The whole environment during the control process mainly consists of three parts: the control system on the field, the electromagnetic propagation environment, and the digital twin on the EIH for system estimation and control command generation. Accordingly, the comprehensive representation of the environment' state at the $t$-th control interval, denoted by $\mathbf{s}_t$, can be formulated as
\begin{equation}
	\mathbf{s}_t = \left\{\mathbf{x}_t, \mathbf{h}_t, \hat{\mathbf{x}}_{t|t-1}, \mathbf{P}_{t|t-1}\right\},
\end{equation}
where $\mathbf{x}_t$ represents the state of the control system at cycle $t$, $\mathbf{h}_t = \left\{h_{k,t}\right\} $ denotes the channel gain that characterizes the communication channel. In addition, the state of the digital twin is determined by the prior estimate $\hat{\mathbf{x}}_{t|t-1}$ of the Kalman filter and its error covariance matrix $\mathbf{P}_{t|t-1}$. 

\subsubsection{Observation Space}
From a practical implementation perspective, the DRL agent cannot access the complete system state $\mathbf{s}_t$. Specifically, the agent is deployed on the EIH. At the beginning of each control, the EIH acquires the channel state information through pilot-based channel estimation. The DRL agent then determines the bandwidth allocation based on the channel conditions and the state of the digital twin. After that, the EIH transmits the bandwidth allocation results to the sensors to guide their transmissions. Therefore, the accurate physical state $\mathbf{x}_t$ is not available for the agent. Consequently, the decision-making process is based on an observation space that encapsulates all information available at the EIH. The observation at the $t$-th control interval can be expressed as
\begin{equation}\label{observation}
	\mathbf{o}_t = \left\{\mathbf{h}_t, \hat{\mathbf{x}}_{t|t-1}, \mathbf{P}_{t|t-1}\right\}.
\end{equation}

It is worth noting that the terms $\hat{\mathbf{x}}_{t|t-1}$ and $\mathbf{P}_{t|t-1}$ provide a comprehensive estimate for $\mathbf{x}_t$ based on the entire history of past measurements and actions, thereby approximately satisfying the Markov property required for reinforcement learning and ensuring the efficacy of our proposed DRL approach.

\subsubsection{Action Space}
Based on the observation $\mathbf{o}_t$, the DRL agent designs the bandwidth allocation of each sensor. The raw output of the agent's policy network is a continuous action vector as
\begin{equation}
	\mathbf{a}_t = \left[a_{1,t}, a_{2,t}, \cdots, a_{K,t}\right].
\end{equation}
To enhance the stability of the training process, the components $a_{k,t} $ are constrained to the interval $\left( -1,1\right)$.  This raw action is then mapped to the actual bandwidth allocation $B_{k,t} $ for sensor $k$ as
\begin{equation}
B_{k,t} = \frac{a_{k,t}+1}{\sum_{i=1}^{K}\left( a_{i,t}+1\right) }B_{\text{max}}.
\end{equation}

\subsubsection{Reward Function and Optimization Objective}
The objective of the DRL agent is to learn a policy that minimizes the long-term LQR control cost defined in \eqref{LQR}. To align with the standard DRL framework of reward maximization, we define the instantaneous reward at cycle $t$ as
\begin{equation}\label{rt}
	r_t = -\left( \mathbf{x}_{t}^\text{T}\mathbf{Q}\mathbf{x}_{t} +\mathbf{u}_{t}^\text{T}\mathbf{R}\mathbf{u}_{t}\right).
\end{equation}
{The reward function in \eqref{rt} is directly derived from the objective function in \eqref{P1a}. Specifically, the term $ \mathbf{x}_{t}^\text{T}\mathbf{Q}\mathbf{x}_{t} +\mathbf{u}_{t}^\text{T}\mathbf{R}\mathbf{u}_{t}$ represents the instantaneous LQR cost at control cycle $t$. Since the standard DRL framework aims to maximize the accumulated reward, we define the reward as the negative instantaneous LQR cost. Therefore, maximizing the long-term accumulated reward encourages the agent to minimize the long-term LQR cost, which is consistent with the objective in \eqref{P1a}.}

Although the reward at each control cycle is defined as the instantaneous LQR cost, the bandwidth allocation action has a long-term impact on the control performance as it can affect the estimation accuracy and the system state. It should be noted that the DRL agent does not update the policy based solely on the one-step reward. Instead, it is trained to maximize the expected discounted cumulative return, i.e.,
\begin{equation}
	\max_{\pi} \quad  \mathbb{E}\left[\sum_{i=0}^{\infty}\gamma^i r_{t+i}\right],
\end{equation}
where $\gamma\in \left( 0,1\right) $ is the discount factor that balances the long-term and short-term rewards. By setting $\gamma$ close to one, the DRL agent learns a policy that accounts for the entire sequence of future rewards, which is consistent with the objective of minimizing the infinite-horizon LQR in \eqref{LQR}.

\subsection{PPO-Based Reinforcement Learning Method}
Our framework assumes continuous observation and action spaces so that discrete-action value-based methods (e.g., Q-learning or deep Q-network) may not be used. In addition, a stable training process is crucial in our control-oriented framework: a poor allocation policy can destabilize the closed loop, leading to an extremely large LQR cost and high return variance. PPO offers a favorable stability–complexity trade-off compared with the alternatives~\cite{PPO}, e.g., deep deterministic policy gradient (DDPG) and, soft actor-critic (SAC), and trust region policy optimization (TRPO). Therefore, we choose the PPO algorithm to train the DRL agent.

The key innovation of PPO to improve the training stability is that it introduces a clipped surrogate objective function to limit excessive changes of the policy during the training process. The clip loss function can be formulated as 
\begin{align}
	L^{\text{CLIP}} = \mathbb{E} \left[ \min \left( r_t(\theta) A_t, \text{clip}(r_t(\theta), 1 - \epsilon, 1 + \epsilon) A_t \right) \right],
\end{align}
where $\theta$ represents the parameters of the policy network, and $r_t(\theta) = \frac{\pi_\theta(\mathbf{a}_t | \mathbf{o}_t)}{\pi_{\theta_\text{old}}(\mathbf{a}_t | \mathbf{o}_t)}$ is the probability ratio between the new policy and the old policy, where $\pi_\theta(\mathbf{a}_t | \mathbf{o}_t)$ and $\pi_{\theta_\text{old}}(\mathbf{a}_t | \mathbf{o}_t)$ denote the probability of the action $\mathbf{a}_t$ generated by the new policy and the old policy, respectively. The function $\text{clip}(r_t(\theta), 1 - \epsilon, 1 + \epsilon)$ denotes the clip operation that restricts the ratio within the interval $\left[ 1 - \epsilon, 1 + \epsilon\right]$, where $\epsilon$ is a hyperparameter that defines the clipping range. $A_t$ is the advantage function, which estimates how much better a specific action $\mathbf{a}_t$ is compared to the average action in observation $\mathbf{o}_t$. The advantage function $A_t$ can be estimated based on the generalized advantage estimation (GAE) method, as
\begin{align}\label{GAE}
A_t = \delta_t + (\gamma \lambda) \delta_{t+1} + \dots + (\gamma \lambda)^{T-t-1} \delta_{T-1},
\end{align}
where $\delta_t = r_t + \gamma V(\mathbf{o}_{t+1}) - V(\mathbf{o}_t)$ denotes the temporal difference (TD) error which measures the difference between the current value prediction and the estimate based on the next-step observation, $V(\mathbf{o}_t)$ is the value function estimated by the critic network, and $\lambda$ is the GAE hyperparameter that balances the bias and variance of the estimate. By computing $A_t$ as the discounted sum of multi-step TD errors ($\delta_{t+i}$), GAE effectively aggregates the influence of the entire future sequence of rewards and value predictions into a single advantage signal. The clip loss function aims to improve the action value while avoiding large updates in the policy, thus ensuring the stability during the training process.

The PPO algorithm also includes the value loss for the critic network, which is computed as the mean squared error (MSE) between the predicted value of the critic network and the actual return
\begin{align}
	L^{\text{VF}} = \frac{1}{2} \left[ (V(\mathbf{o}_t) - R_t)^2 \right],
\end{align}
where $R_t$ is the discounted sum of rewards.

In addition, to encourage exploration, PPO utilizes the entropy term that penalizes deterministic policies. The entropy loss function can be calculated as
\begin{equation}
	L^{\text{entropy}} = - \pi_\theta(\mathbf{a}_t | \mathbf{o}_t) \log \pi_\theta(\mathbf{a}_t | \mathbf{o}_t).
\end{equation}

The overall loss function is the weighted sum of the above three loss functions, which can be rewritten as 
\begin{equation}\label{loss}
L^{\text{PPO}} = \mathbb{E}_t\left( L^{\text{CLIP}} - c_1 L^{\text{VF}} + c_2 L^{\text{entropy}} \right),
\end{equation}
where $c_1$ and $c_2$ are parameters that control the relative importance of the value function loss and entropy loss. Based on the above loss function, the DRL network parameters can be updated based on the gradient descent method.

\begin{figure*} [t]
	\centering
	\includegraphics[width=0.65\linewidth]{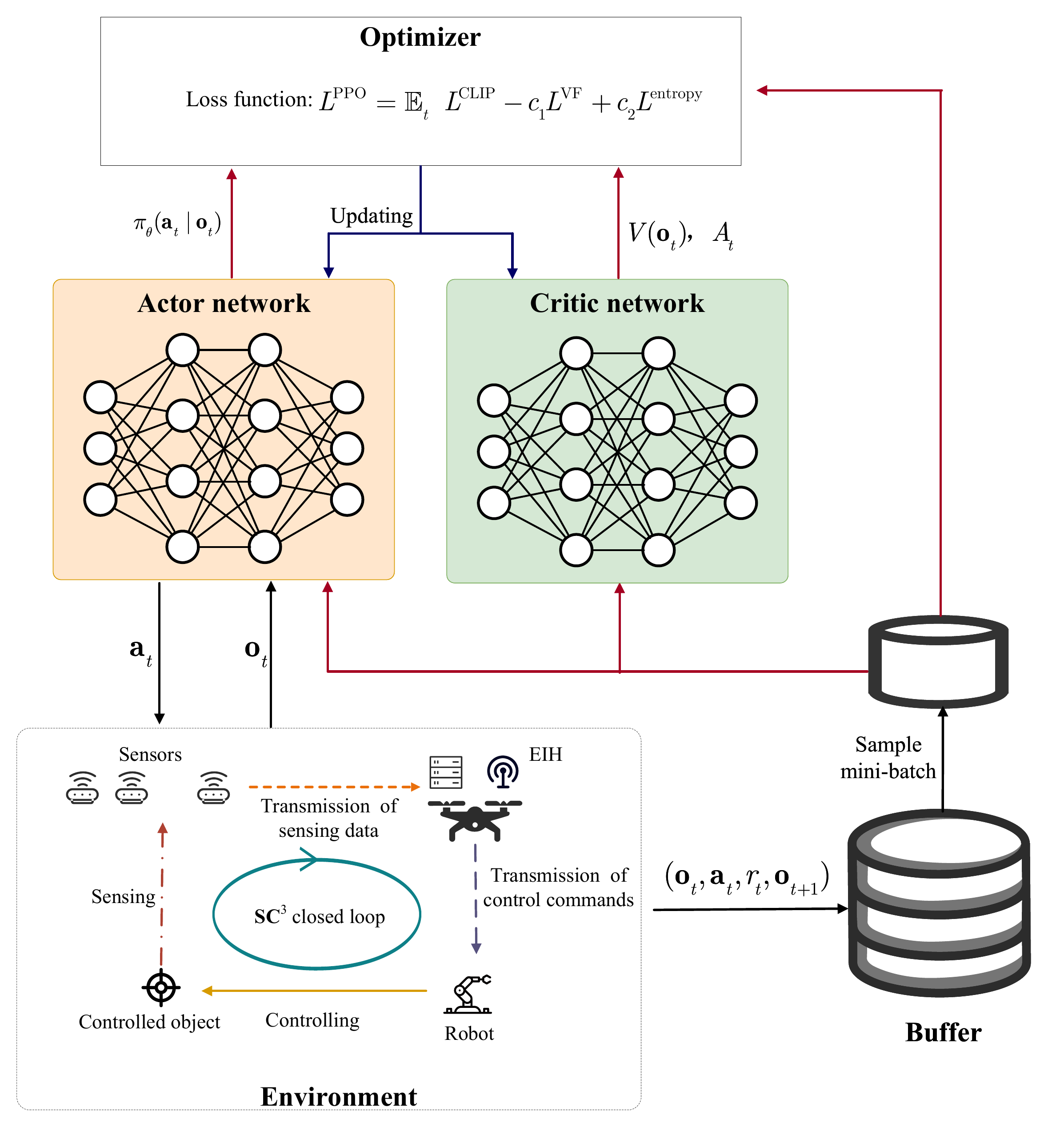}
	\caption{Illustration of the proposed PPO-based bandwidth allocation framework.}
	\label{fig:structure}
\end{figure*}

To illustrate our proposed PPO-based bandwidth allocation framework, Fig. \ref{fig:structure} describes the training and the PPO-based DRL framework, and the training algorithm is also given in \textbf{Algorithm \ref{Algo1}}.  

\subsection{Adaptations of PPO for the considered problem}

{To make the PPO algorithm suitable for the considered bandwidth allocation problem, we introduce several problem-specific designs. First, the observation of the agent consists of heterogeneous information, including the channel state, the prior state estimate, and the error covariance matrix. These components have different physical meanings and dimensions. Therefore, instead of directly concatenating all input features, we design a three-path feature extraction network to separately process different parts of the observation. The extracted features are then concatenated and further processed by a residual block, which helps the actor and critic networks learn useful representations from the communication and control-related information.}

{Second, we introduce a training stabilization mechanism for the closed-loop control scenario. During exploration, the DRL agent may select bandwidth allocation actions that lead to poor state estimation and even control system instability, resulting in extremely large LQR costs and highly negative rewards. Such trajectories can introduce high-variance gradients and destabilize the PPO training process. Therefore, we apply early termination for severely unstable trajectories, so as to keep the reward magnitude in a suitable numerical range.} Specifically, we define an LQR cost threshold $l_{\text{th}}$. If the instantaneous cost, i.e., $-r_t$, exceeds this threshold for a predefined number of consecutive control cycles (e.g., three), the system is deemed unstable. In such cases, the current episode will be terminated, and the environment is reset to its initial state to start another episode.

\begin{algorithm}[t]\label{Algo1}
\caption{PPO-Based Training Algorithm.}
\SetKwInOut{Input}{Input}\SetKwInOut{Output}{Output}\SetKwInOut{Initialize}{Initialization}

\Input {System parameters: System matrix $\mathbf{\Phi}$, input matrix  $\mathbf{\Gamma}$, bandwidth constraint $B_{\text{max}}$, etc. DRL Hyperparameters: discount factor $\gamma$, GAE parameter $\lambda$, clipping threshold $\epsilon$, loss coefficients $c_1, c_2$, training epochs $E$, and instability cost threshold $l_{\text{th}}$.}
Initialize the actor network and critic network\;
\For{iteration number = $1, 2, \cdots$}{
	\tcp{Phase 1: Data Collection}
Clear the rollout buffer\;
\While{the buffer is not full}{
	Reset the environment and obtain initial observation $\mathbf{o}_0$.\;
	\For{$t = 0, 1, 2, \dots, t_{\text{max}}-1$}{
		Generate action $\mathbf{a}_t$ from the actor network based on the observation $\mathbf{o}_t$\;
		Execute action $\mathbf{a}_t$, observe reward $r_t$ and next observation $\mathbf{o}_{t+1}$\;
		Store the transition $(\mathbf{o}_t, \mathbf{a}_t, r_t, \mathbf{o}_{t+1})$ in buffer\;
		\If{The control system is unstable}{break.\tcp{Terminate episode}}
	}
}
\tcp{Phase 2: Compute Advantages and Returns}
\For{each trajectory in the buffer}{Compute advantage estimates $A_t$ based on \eqref{GAE}\;
	Compute the discounted sum of rewards $R_t$;}
\tcp{Phase 3: Update Network Parameters}
\For{Epoch number $= 1, 2, \cdots, E$}{
Shuffle the data in the buffer and split them into mini-batches\;
\For{each mini-batch}{
	Calculate the total loss function according to \eqref{loss}\;
	Update network parameters using a gradient-based optimizer.
}
}}
\Output  {The trained DRL agent for bandwidth allocation.}
\end{algorithm}

\subsection{{Practical Implementation of the Proposed Framework}}
{In practice, the proposed framework can be implemented through two phases, i.e., offline training and online deployment. In the offline training phase, the DRL agent is trained based on \textbf{Algorithm \ref{Algo1}}, where a simulated environment is constructed according to the system dynamics, sensing model, and channel model. After training, only the actor network needs to be deployed at the EIH for online bandwidth allocation. Specifically, at the beginning of each control cycle, the EIH estimates the channel state information through pilot-based channel estimation. Meanwhile, the EIH maintains the digital twin of the control system and updates the prior state estimate and the corresponding error covariance matrix according to \eqref{update1} and \eqref{update2}. Next, the observation defined in \eqref{observation} is fed into the trained actor network to output the bandwidth allocation action. The EIH then transmits the bandwidth allocation results to the sensors, and the sensors transmit their sensing data using the allocated bandwidths. After receiving the sensing data, the EIH updates the state estimate based on the Kalman filter, calculates the control input according to the LQR controller, and sends the control command to the robot. It should be noted that the online deployment phase does not require DRL network parameter updates. The computation complexity mainly comes from the Kalman filtering and one forward propagation of the actor network in each control cycle, which can be efficiently executed at the EIH. If the environment or control task changes significantly, the policy can be periodically retrained or fine-tuned offline and then redeployed at the EIH.}

\section{Simulation Results and Discussions}
\label{sec_simulation}
In this section, we provide simulation results to show the performance of the proposed DRL-based bandwidth allocation scheme. We first present simulation setting, and then discuss the simulation results.
\subsection{Simulation Setting}
Unless specified otherwise, we assume that there are $K=4$ sensors in the control system, all of which are randomly distributed in a circular area with a radius of $5000$ m. The UAV, to elevate the EIH, is located at the center of the circular area with a height of $100$ m. The communication environmental parameters are set as $\eta_{\text{LOS}} = 0.1$, $\eta_{\text{NLOS}} = 21$, $a = 4.88$ and $b = 0.43$ corresponding to the suburban scenario~\cite{channel1}. Unless specified otherwise, the bandwidth constraint is set as $B_{\text{max}}=50$ kHz, the transmit time is set as $T = 1$ ms, and the transmit power is $p_{k,t}=0.1$ W. Other communication parameters are: $N_0 = -174$ dBm/Hz, $c = 3e8$ m/s, $f = 2$ GHz, and $\rho = 0.005$.

\begin{table}[!t]
	\centering
	\caption{{PPO hyperparameters and training settings.}}
	\label{tab:ppo_training_settings}
	{\begin{tabular}{p{3.0cm}p{4.0cm}}\toprule
		\textbf{Parameter} & \textbf{Value}\\\midrule
		Total training steps & $4\times 10^{5}$ \\\hline
		Learning rate & $2\times 10^{-4}$ \\\hline
		Rollout length & 2048 \\\hline
		Batch size & 256 \\\hline
		Discount factor $\gamma$ & 0.99 \\\hline
		GAE parameter $\lambda$ & 0.95 \\\hline
		PPO clipping range & 0.2 \\\hline
		Entropy coefficient & 0 \\\hline
		Value-function coefficient & 0.5 \\\hline
		Maximum gradient norm & 0.5 
		\\\bottomrule
	\end{tabular}}
\end{table}

For the control system, we set $n = 4$ and $m = 4$. The parameter matrices are set as \cite{distortion_noise}
\begin{align}
	&\mathbf{\Phi} =
	\begin{bmatrix}
		0.12 & 0.63 & -0.52 & 0.33 \\
		0.26 & -1.28 & 1.57 & 1.13 \\
		-1.77 & -0.30 & 0.77 & 0.25 \\
		-0.16 & 0.20 & -0.58 & 0.56
	\end{bmatrix}, \label{simuA}\\
	&
	\mathbf{\Gamma} =
	\begin{bmatrix}
		0.66 & -0.58 & 0.03 & -0.20 \\
		2.61 & -0.91 & 0.87 & -0.07 \\
		-0.64 & -1.12 & -0.19 & 0.61 \\
		0.93 & 0.58 & -1.18 & -1.21
	\end{bmatrix},
	\\
	&\mathbf{V} =
	\begin{bmatrix}
		4.94 & -0.10 & 1.29 & 0.35 \\
		-0.10&5.55 & 2.07 & 0.31 \\
		1.29&2.07&2.02 & 1.43 \\
		0.35 &0.31 &1.43 & 3.10
	\end{bmatrix},
\end{align}
The covariance matrix $\mathbf{P}_1$ of the initial state $\mathbf{x}_1$ is generated randomly by calculating $\mathbf{P}_1 = \mathbf{H}^\text{T}\mathbf{H}$, where $ \mathbf{H}\in \mathbb{R}^{n\times n}$, and the entries of $\mathbf{H}$ are selected randomly according to the Gaussian distribution $\mathcal{N}(0,1)$. The LQR weight matrices are set as $\mathbf{Q} = 0.01\times I_n$ and $\mathbf{R} = 0.01\times I_m$ , where $I_n$ denotes the identity matrix of size $n$. The sensing parameters are set as $	\mathbf{C} = I_K $, and $\mathbf{\sigma}_{\text{w},k}^2 = 0.1 $ unless specified otherwise.

\begin{figure} [t]
	\centering
	\includegraphics[width=\linewidth]{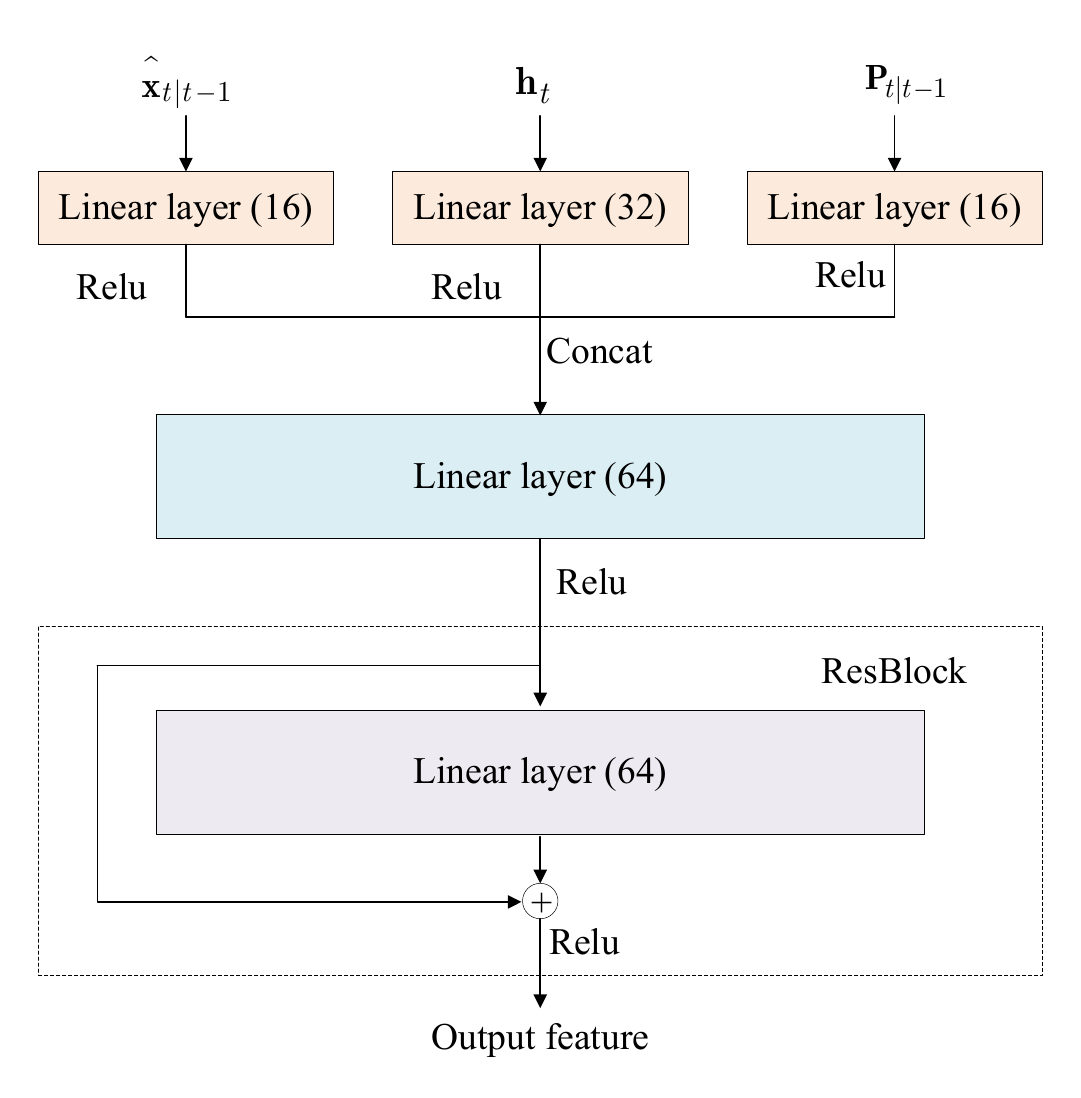}
	\caption{Structure of the feature extraction network.}
	\label{fig:network}
\end{figure} 

For deep reinforcement learning, we conducted training based on Stable-Baselines3 library~\cite{sb3}, an open-source Python library for reinforcement learning. The architecture of the feature extraction network is illustrated in Fig. \ref{fig:network}. The model employs a shared feature extractor with a three-path parallel architecture to process the heterogeneous input modalities. The outputs from these paths are concatenated and passed through another 64-unit linear layer, followed by a residual block (ResBlock) to enhance feature representation capability. The extracted features are then fed into both the actor network and critic network to estimate the action and advantage. The actor network consists of fully connected layers with output dimensions 64, 64, and 4, while the critic network comprises three fully connected layers with output dimensions of 64, 64, and 1. Both networks utilize tanh as the activation function. In the training process, the maximum steps of each episode are $40$, i.e., the control system executes at most $40$ control cycles before resetting. The main training hyperparameters are shown in TABLE \ref{tab:ppo_training_settings}, all other hyper-parameters of the PPO algorithm follow the default settings of stable-baselines3.

\begin{figure} [t]
	\centering
	\includegraphics[width=\linewidth]{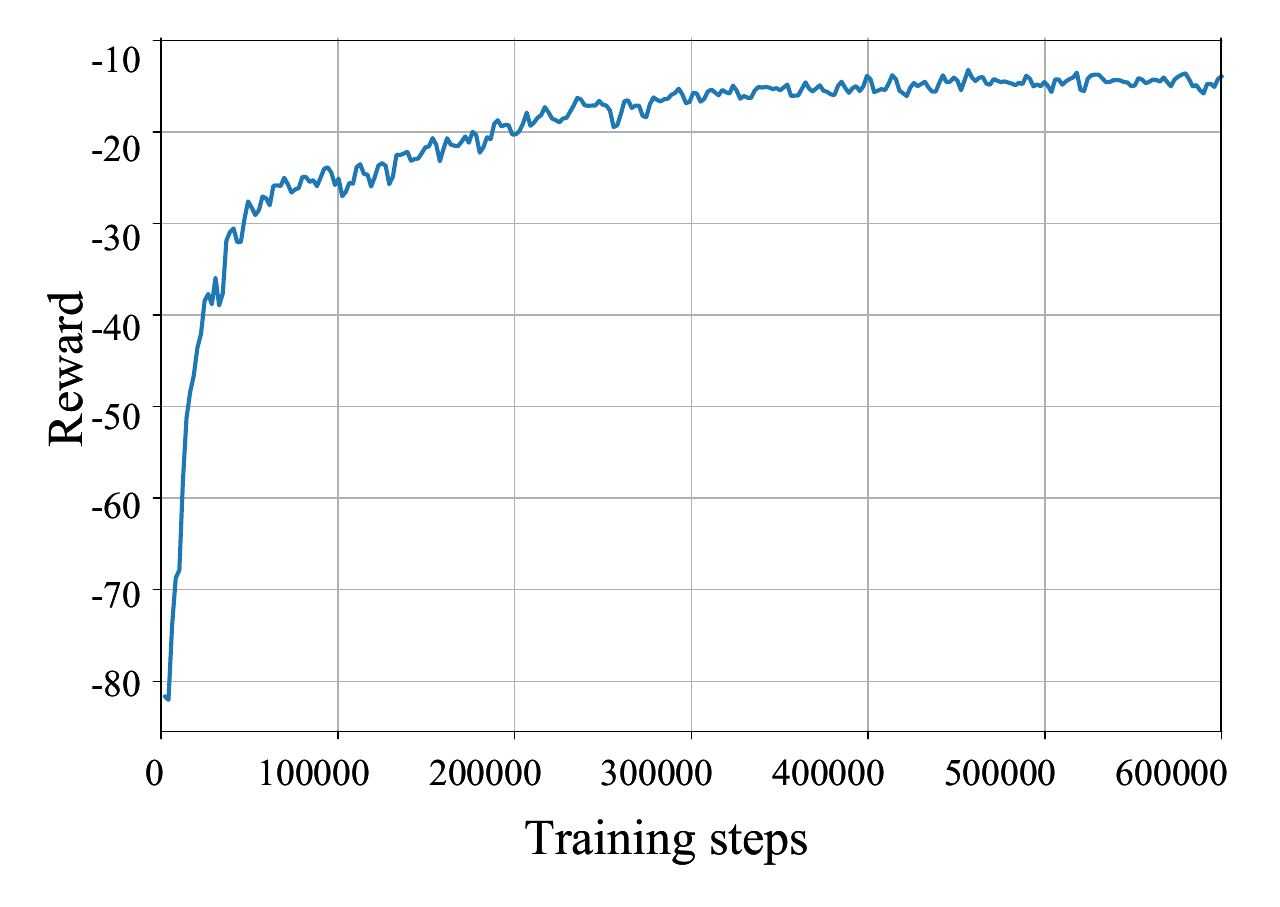}
	\caption{Convergence performance of the proposed DRL-based bandwidth allocation scheme.}
	\label{fig:simu1}
\end{figure}

\subsection{Convergence Performance}
In Fig. \ref{fig:simu1}, we show the convergence of the proposed DRL-based bandwidth allocation scheme, where the bandwidth constraint is set as $50$ kHz. To enhance visual clarity, the rewards displayed in the figure are averaged over $100$ episodes. From the figure, it is shown that the reward increases rapidly during the first $100000$ training steps. After approximately $400000$ training steps, the reward gradually stabilizes, showing the convergence of the proposed DRL-based scheme.

\begin{figure} [t]
	\centering
	\includegraphics[width=\linewidth]{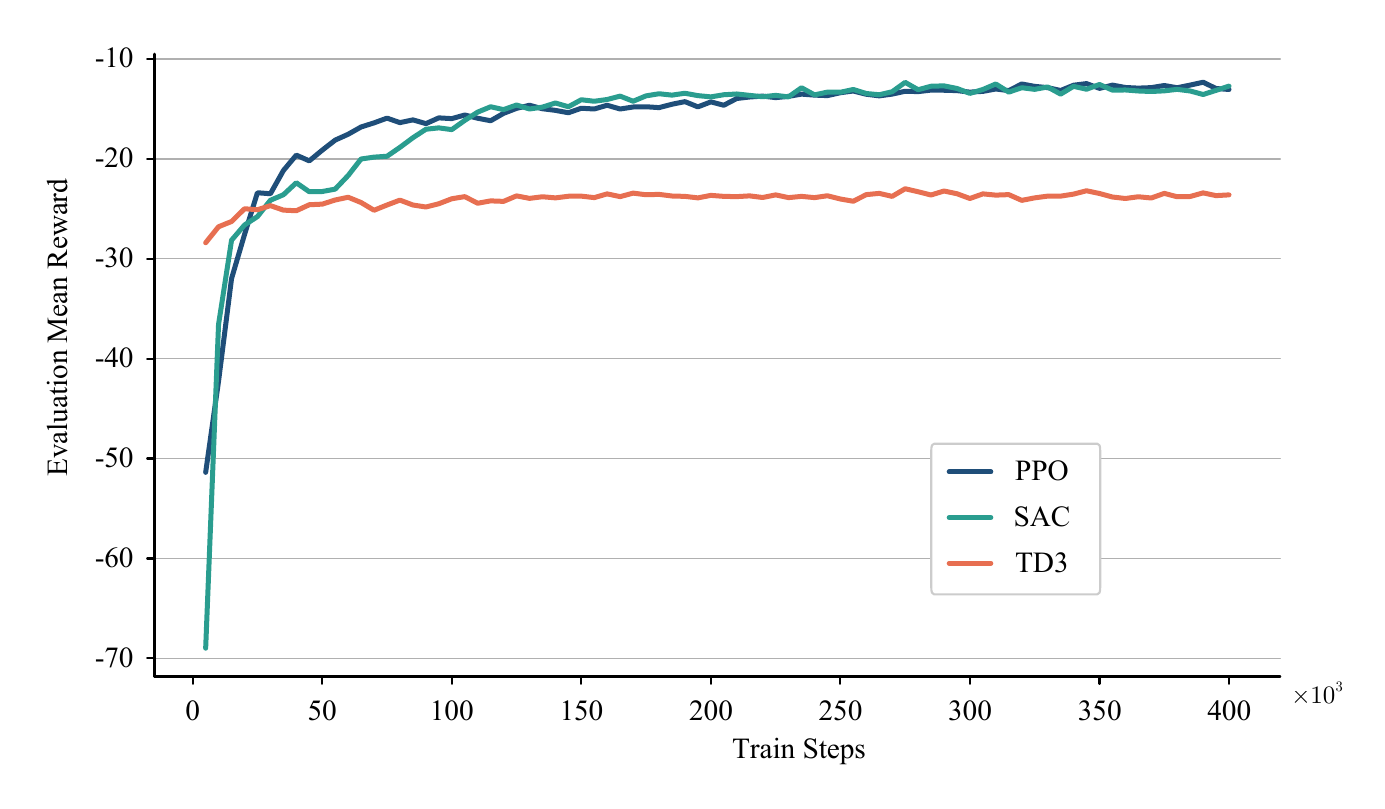}
	\caption{Training performance of representative DRL algorithms.}
	\label{fig:simu_algorithm_training_curve}
\end{figure} 

{We further compare the training performance of PPO with two representative continuous-control DRL algorithms, i.e., SAC and TD3, where the bandwidth constraint is set as $B_{\text{max}} = 40$ kHz. As shown in Fig. \ref{fig:simu_algorithm_training_curve}, TD3 converges to a significantly lower reward, indicating that it is less effective for the considered closed-loop bandwidth allocation problem. PPO achieves a fast improvement in the early training stage and converges to a stable reward after training. SAC can finally achieve a similar reward, but its early-stage training is slower in this setting. In addition, in our setting, PPO requires less training time than SAC. Therefore, we choose PPO as the main algorithm to train the DRL agent.}

\subsection{Control Performance Evaluation}
\begin{figure} [t]
	\centering
	\includegraphics[width=\linewidth]{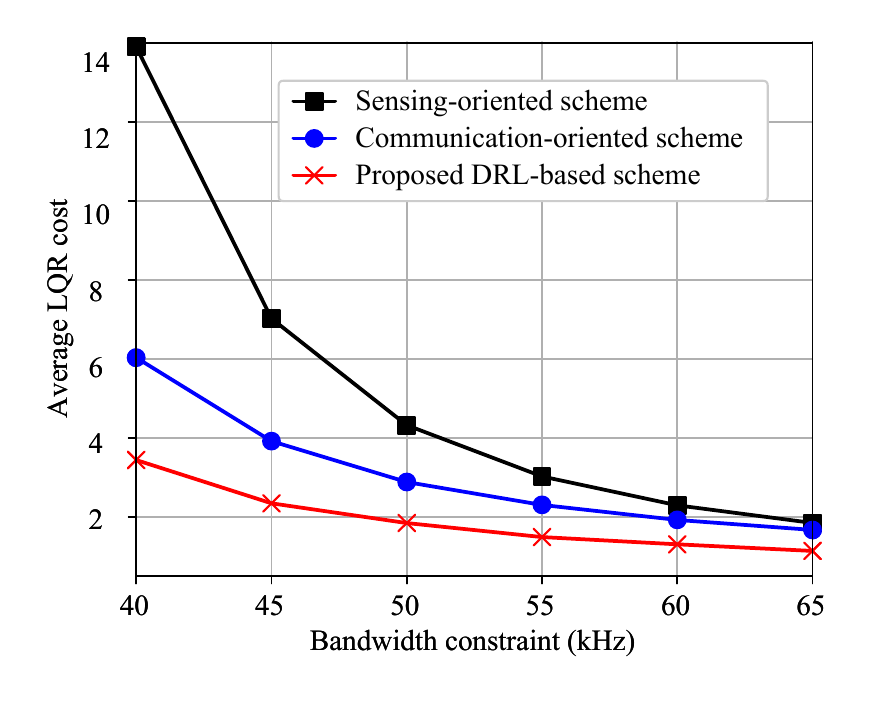}
	\caption{Averaged LQR cost by different schemes varying with the bandwidth constraint.}
	\label{fig:simu_LQR_vs_B}
\end{figure} 

We show the LQR cost per cycle under different bandwidth constraints in Fig. \ref{fig:simu_LQR_vs_B}. Due to the stochastic nature of the environments, we conduct $1000$ simulations and report the averaged value of the LQR costs in the figure. In order to evaluate the performance of the proposed DRL-based scheme, we compare it with two benchmark schemes: (i) the sensing-oriented scheme in \cite{kalman1}, where the sensing performance is optimized by allocating more bandwidth to the sensor with higher sensing signal-noise-ratio (SNR), and (ii) the communication-oriented scheme, which allocates bandwidth to maximize the sum communication rates of the sensor-to-UAV links. The results demonstrate the superiority of our proposed method, as the proposed DRL-based scheme achieves the lowest LQR cost among the three schemes across the entire range of bandwidth constraints. In addition, the LQR cost decreases with the increasing of bandwidth constraint, indicating that enhancing the communication capability is an efficient way to improve the overall performance of the control system.

\begin{figure} [t]
	\centering
	\includegraphics[width=\linewidth]{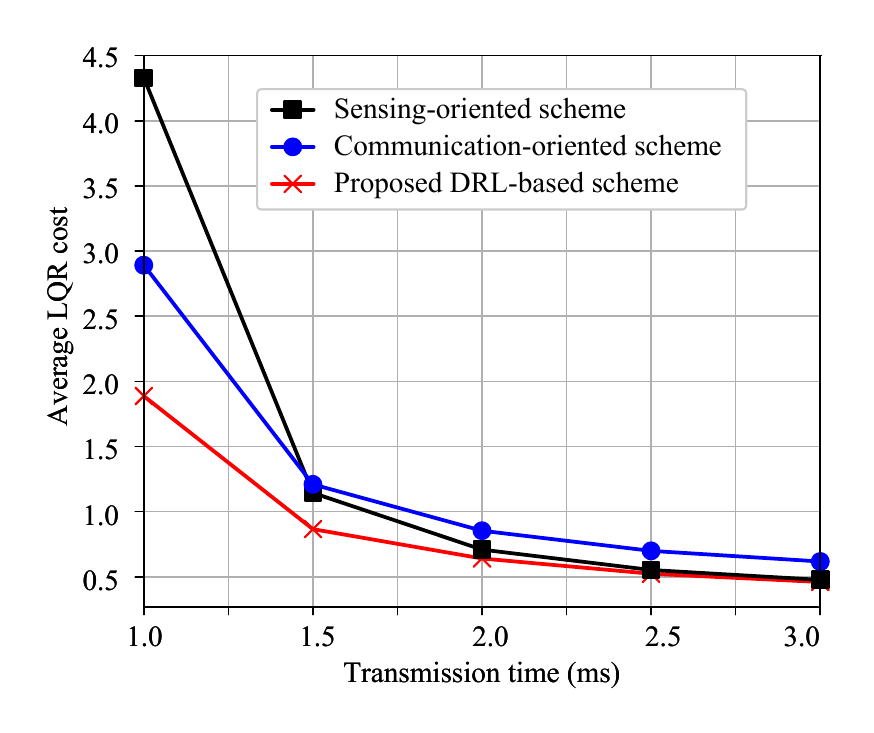}
	\caption{Averaged LQR cost by different schemes varying with the transmission time.}
	\label{fig:simu_LQR_vs_T}
\end{figure} 

\begin{figure} [t]
	\centering
	\includegraphics[width=\linewidth]{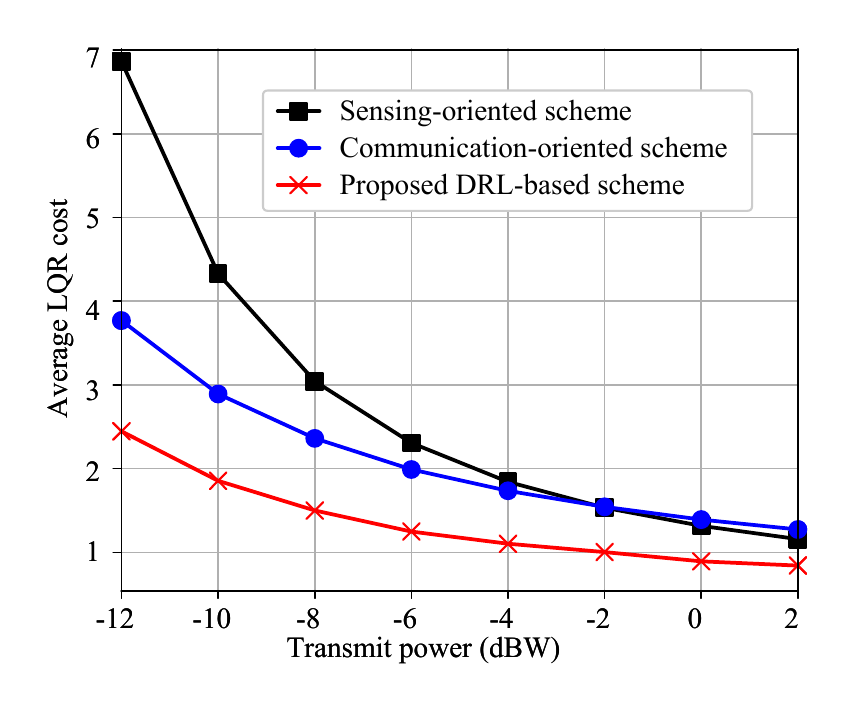}
	\caption{Averaged LQR cost by different schemes varying with the transmit power.}
	\label{fig:simu_LQR_vs_p}
\end{figure} 

In Fig. \ref{fig:simu_LQR_vs_T} and Fig. \ref{fig:simu_LQR_vs_p}, we show the LQR cost with transmission time and transmit power, respectively, where the bandwidth constraint is set as $B_{\text{max}}=50$ kHz. Similar to our previous findings, the proposed DRL-based scheme outperforms the benchmark schemes across all conditions, achieving the lowest LQR cost. The performance advantage of the DRL-based scheme is particularly significant in resource-scarce regime  (i.e., at lower transmission times and power levels). In addition, when $T<1.5$ ms or $p_{k,t}<-2$ dBW, the communication-oriented scheme yields a lower LQR cost than the {sensing-oriented} allocation scheme. However, as the available resources increase, the {sensing-oriented} allocation scheme eventually outperforms the communication-oriented scheme. This indicates that, when the communication resources are abundant, it is more effective to optimize the sensing performance, rather than the communication performance. Additionally, we observe that the average LQR cost decreases with more available communication resources. The reason is that more sensing information can be transmitted with more resources, which facilitates more accurate state estimation and consequently improves the closed-loop control performance.

\begin{figure} [t]
	\centering
	\includegraphics[width=\linewidth]{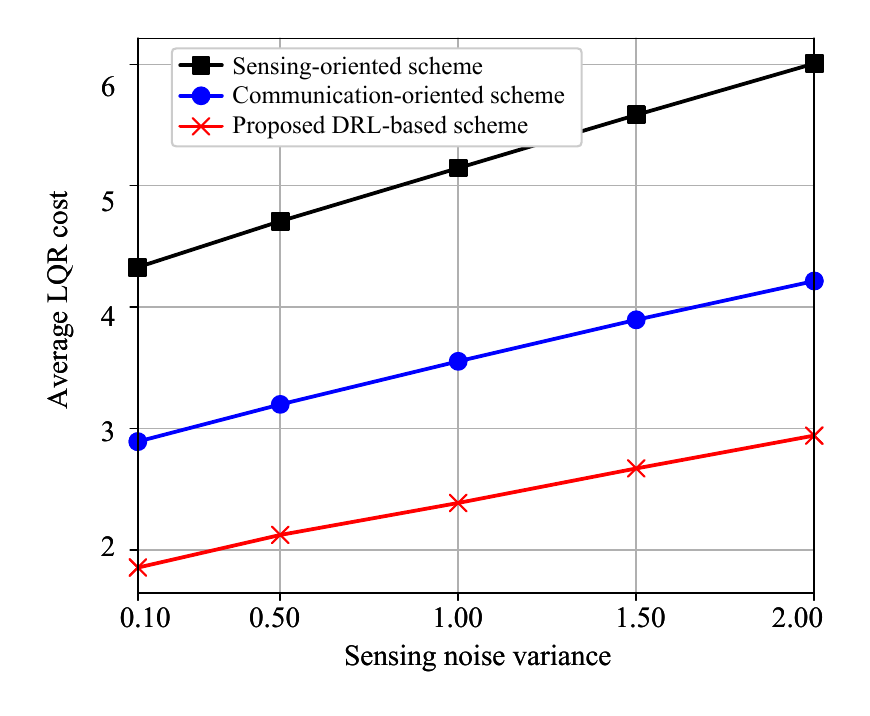}
	\caption{Averaged LQR cost by different schemes varying with the sensing noise variance.}
	\label{fig:simu_LQR_vs_sensor_noise}
\end{figure} 

Fig. \ref{fig:simu_LQR_vs_sensor_noise} evaluates the influence of sensor noise on the average LQR cost for different allocation schemes. In this simulation, we set the sensing variances of each sensor to be the same as $\sigma_{\text{w},k}=\sigma_{\text{w}},\forall k$, and show the relationship between the LQR cost and $\sigma_{\text{w}}$ in Fig. \ref{fig:simu_LQR_vs_sensor_noise}. The results demonstrate that the proposed DRL-based scheme consistently outperforms the other two benchmark schemes, achieving the lowest average LQR cost across the entire range of noise variances tested. In addition, the LQR cost increases monotonically with the sensing noise variance, indicating enhancing the sensing capability by reducing noise variance can improve the control performance.

\begin{figure} [t]
	\centering
	\includegraphics[width=\linewidth]{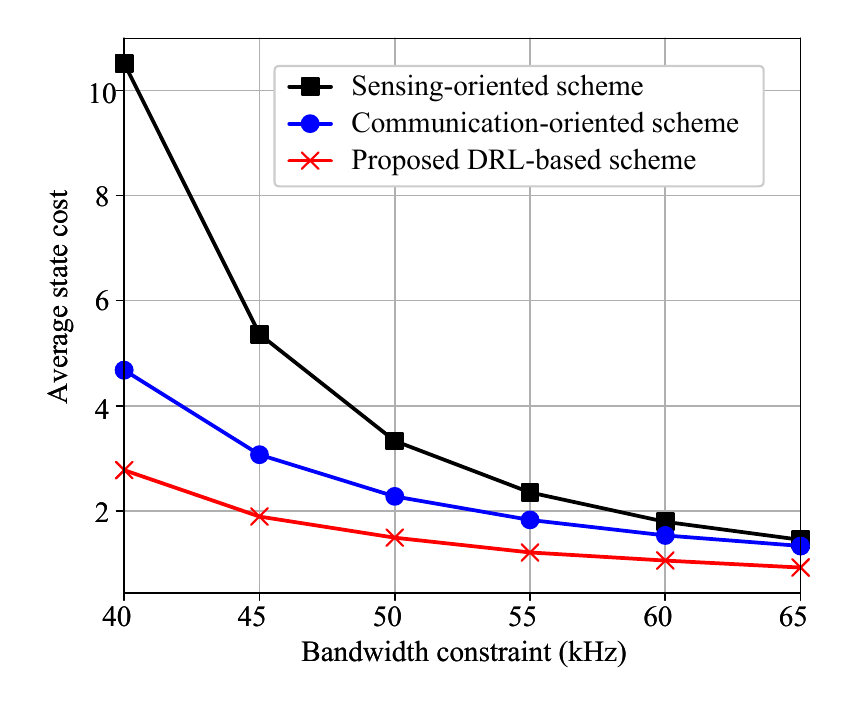}
	\caption{Averaged state cost by different schemes varying with the bandwidth constraint.}
	\label{fig:simu_SC_vs_B}
\end{figure} 

\begin{figure} [t]
	\centering
	\includegraphics[width=\linewidth]{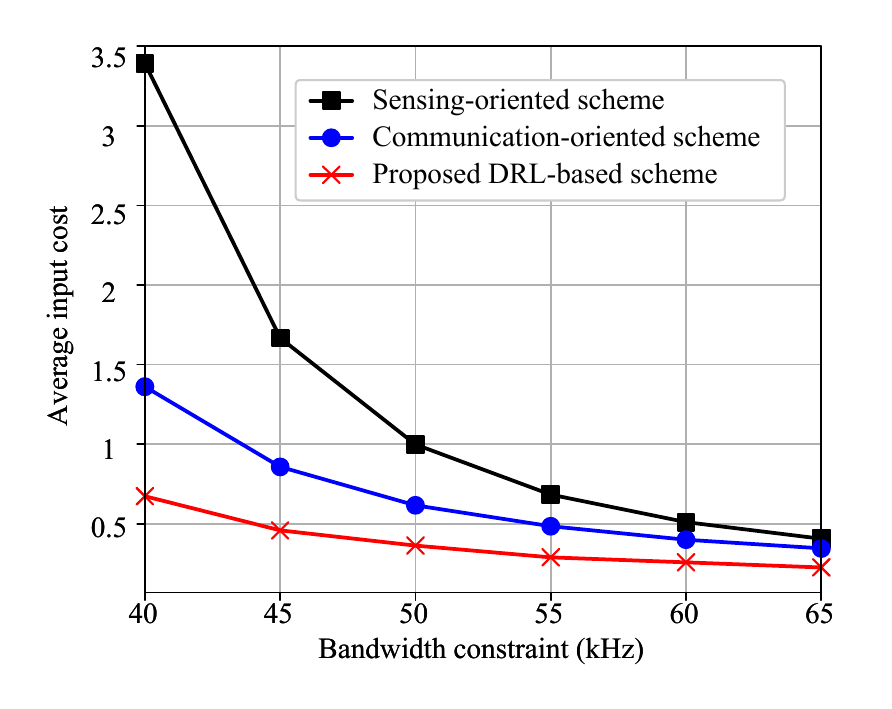}
	\caption{Averaged input cost by different schemes varying with the bandwidth constraint.}
	\label{fig:simu_IC_vs_B}
\end{figure} 

Fig. \ref{fig:simu_SC_vs_B} and Fig. \ref{fig:simu_IC_vs_B} show the average state cost (the term $\mathbf{x}_{t}^\text{T}\mathbf{Q}\mathbf{x}_{t}$ in \eqref{LQR}) and the average input cost (the term $\mathbf{u}_{t}^\text{T}\mathbf{R}\mathbf{u}_{t}$ in \eqref{LQR}) with different bandwidth constraints, respectively. It can be seen that the state cost and input cost have similar trend to the LQR cost shown in Fig. \ref{fig:simu_LQR_vs_B}, indicating that the LQR controller can achieve a tradeoff between the state convergence and input energy consumption. In addition, the proposed DRL-based scheme achieves lower state and input costs than the benchmark schemes. This demonstrates the superiority of our approach in improving the control performance.

\begin{figure} [t]
	\centering
	\includegraphics[width=\linewidth]{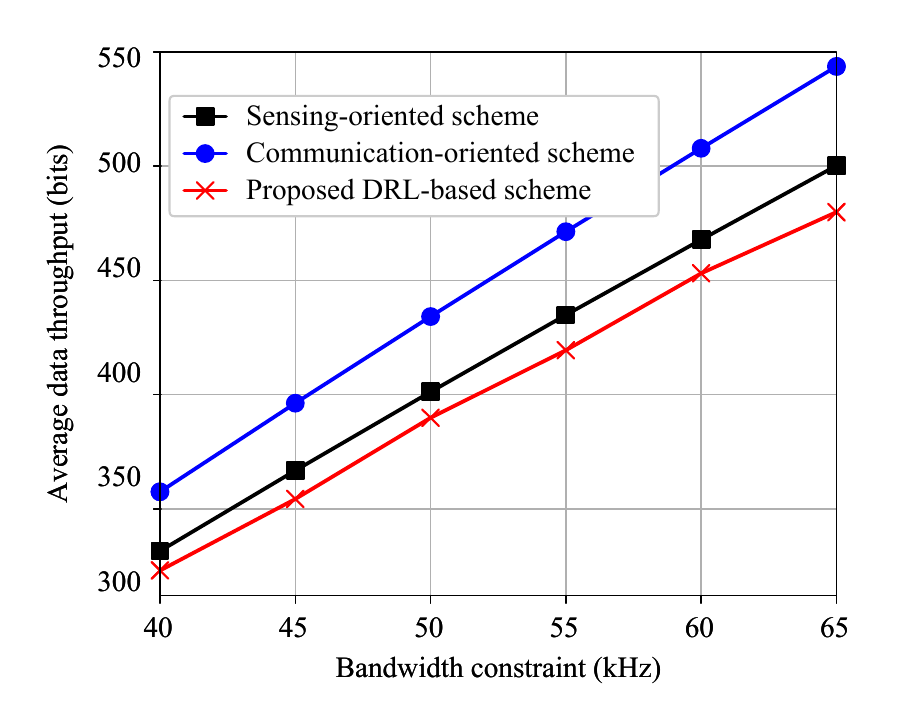}
	\caption{Averaged data throughput by different schemes varying with the bandwidth constraint.}
	\label{fig:simu_DR_vs_B}
\end{figure} 

In Fig. \ref{fig:simu_DR_vs_B}, we compare the average data throughput per control cycle achieved by the three schemes. Intuitively, the communication-oriented scheme yields the highest data throughput among the three schemes. Notably, our proposed scheme results in the lowest sum throughput among the benchmarks. However, the proposed scheme achieves the best control performance among the schemes, as shown in Fig. \ref{fig:simu_LQR_vs_B}. This result indicates that simply maximizing a communication metric, such as sum throughput, is sub-optimal for the overall performance of the $\textbf{SC}^3$ closed loop.

\begin{figure} [t]
	\centering
	\includegraphics[width=\linewidth]{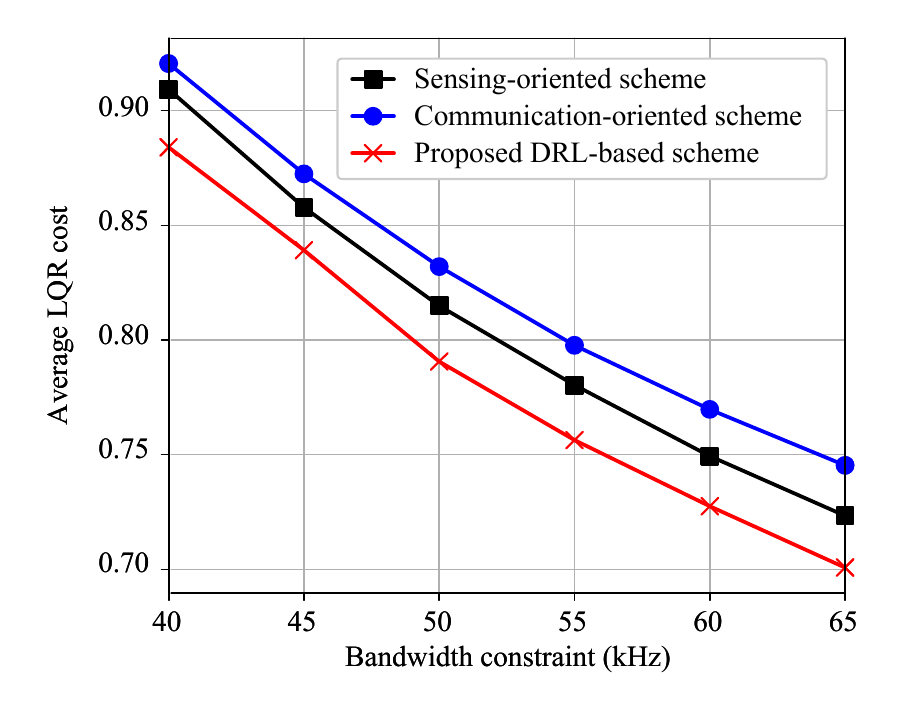}
	\caption{Averaged LQR cost for different schemes with $K=10$.}
	\label{fig:simu_LQR_vs_B_senosr10}
\end{figure} 

Fig. \ref{fig:simu_LQR_vs_B_senosr10} shows the LQR cost with bandwidth constraints when there are $K=10$ sensors in the system. The state dimension is also set as $n=10$. The state transition matrix $\mathbf{\Phi}$ is randomly generated with Gaussian entries, while the control matrix $\mathbf{\Gamma}$ and sensor matrix $\mathbf{C}$ are set as identity matrices. From this figure, we can see that the proposed DRL-based scheme still outperforms the benchmark schemes when the number of sensors increases, demonstrating good scalability and robustness of the proposed approach.

\begin{figure} [t]
	\centering
	\includegraphics[width=\linewidth]{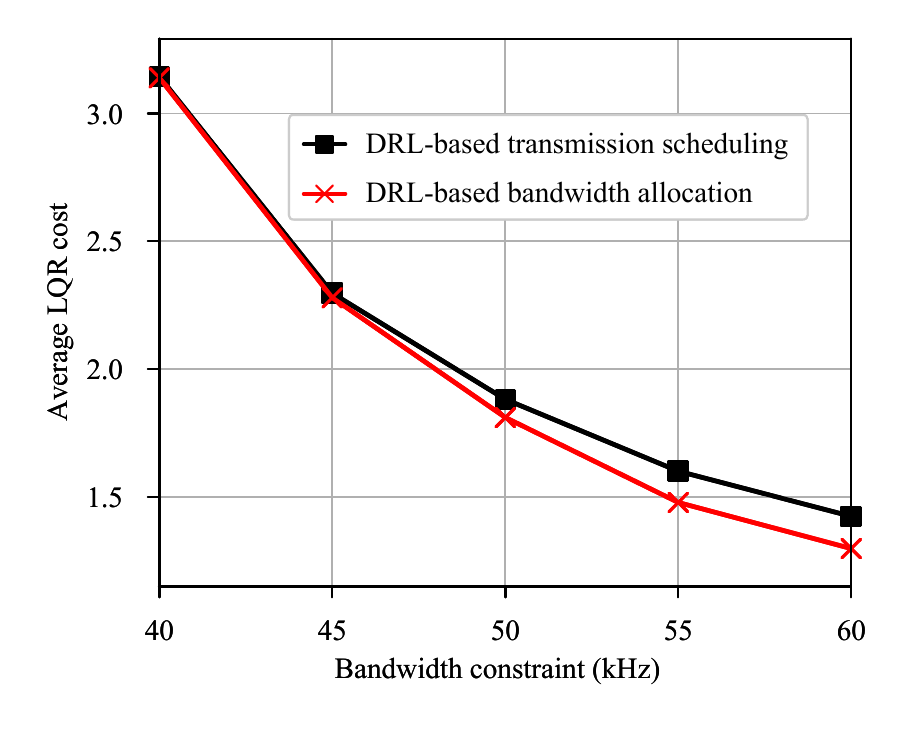}
	\caption{{Averaged LQR cost by two schemes varying with the bandwidth constraint.}}
	\label{fig:simu_LQR_vs_bandwidth1}
\end{figure} 

{Fig. \ref{fig:simu_LQR_vs_bandwidth1} compares the performance of the proposed DRL-based bandwidth allocation scheme with a DRL-based transmission scheduling scheme inspired by \cite{sensor6}. In the transmission scheduling scheme, the DRL agent determines whether each sensor transmits in each control cycle, and the available bandwidth is equally allocated among the active sensors. It can be observed that the proposed DRL-based bandwidth allocation scheme achieves comparable or lower LQR cost than the transmission scheduling scheme, especially when the bandwidth constraint becomes larger. This indicates that directly optimizing continuous bandwidth allocation can more effectively utilize communication resources for improving the closed-loop control performance.}

\subsection{Impact of System Parameters on Bandwidth Allocation Results}
\begin{figure} [t]
	\centering
	\includegraphics[width=\linewidth]{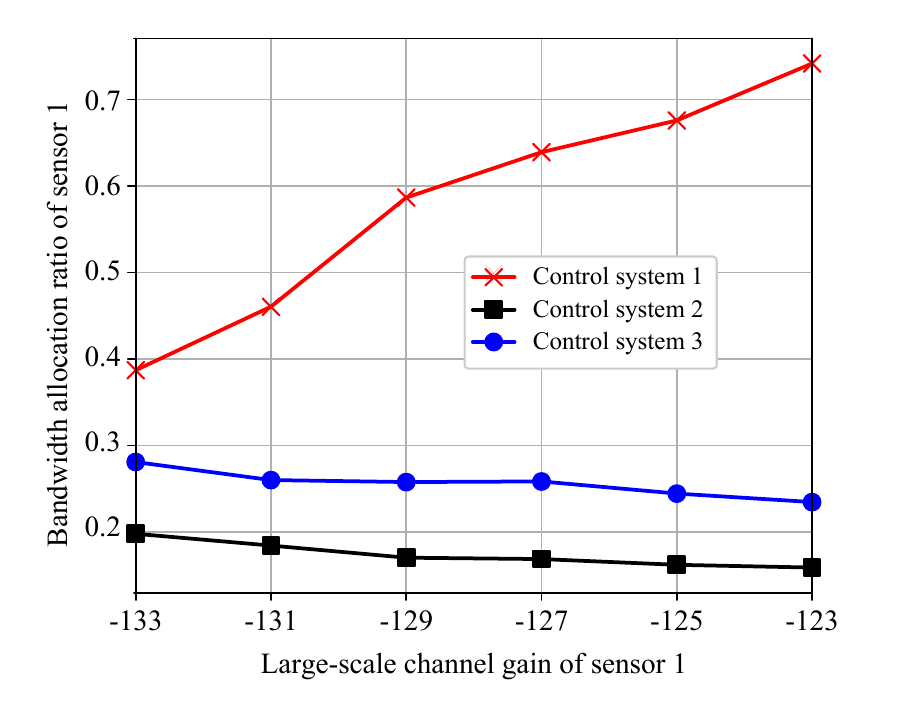}
	\caption{Bandwidth allocation results vs. channel gain under different system dynamics.}
	\label{fig:simu3}
\end{figure} 

In Fig. \ref{fig:simu3}, we show the relationship between the bandwidth allocation results based on the DRL-based scheme and the channel conditions under different control system dynamics. Specifically, we analyze the bandwidth allocation ratio assigned to sensor 1 as its large-scale channel gain (i.e., $g_1$) varies, considering three distinct control systems characterized by different state transition matrices (i.e., $\mathbf{\Phi}$). For each combination of system dynamics and channel gain, a separate DRL agent is trained. To ensure robust results, the average bandwidth allocation ratio over 1000 simulation runs, using a fixed random seed for noise generation, is reported for each setting. We consider three typical system matrices $\mathbf{\Phi}_1$, $\mathbf{\Phi}_2$, and $\mathbf{\Phi}_3$. The matrix $\mathbf{\Phi}_1$ is the same as in \eqref{simuA} , $\mathbf{\Phi}_2$ is set as the diagonal matrix $\text{diag}\left(1.5,1.5,1.5,1.5 \right) $,  representing decoupled subsystems, and $\mathbf{\Phi}_3$ is set as the following block diagonal matrix
\begin{align}
	&\mathbf{\Phi}_3 =
	\begin{bmatrix}
		1.2 & 0.63 & 0 & 0 \\
		0.26 & -1.28 & 0 & 0 \\
		0 & 0 & 1.5 & 0 \\
		0 & 0 & 0 & 1.5
	\end{bmatrix}.
\end{align}
From Fig. \ref{fig:simu3}, we observe that the relationship between the channel gain of a sensor and its allocated bandwidth ratio is highly dependent on the underlying system dynamics. For control system 1, the bandwidth allocation ratio increases with the corresponding channel gain. In contrast, for the second and third systems, the bandwidth allocation ratio approximately decreases with the corresponding channel gain. This phenomenon can be explained as follows. For the first control system, all the components of the system state are coupled closely. The information from a single sensor is valuable not only for its directly observed state but also for estimating other coupled states via the Kalman filter. Consequently, the learned policy tends to allocate more resources to the sensors with better channel conditions, thereby improving the estimate of the entire system state. In contrast, for the second and third control systems which contain multiple decoupled subsystems, the sensor information benefits only its corresponding subsystem. Improving sensor 1’s channel will lower the marginal return of allocating additional bandwidth to it. Therefore, the policy allocates resources to the sensor with weaker channels to achieve a larger overall reduction for the overall control performance. These results indicate that there is not a universal allocation dictating how bandwidth should be allocated based solely on channel gain. In fact, the monotonicity of relationship between the bandwidth allocation ratio and channel gain depends on the control system parameters. {This observation indicates that the bandwidth allocation should not be determined only by the channel quality of sensors in practical deployments. Instead, the coupling structure of the control system states should also be considered. For strongly coupled control systems, allocating more bandwidth to the sensors with better channel quality may benefit the estimation of the whole system state, while for decoupled systems, resources may need to be balanced among different subsystems to avoid performance bottlenecks.}

\begin{figure} [t]
	\centering
	\includegraphics[width=\linewidth]{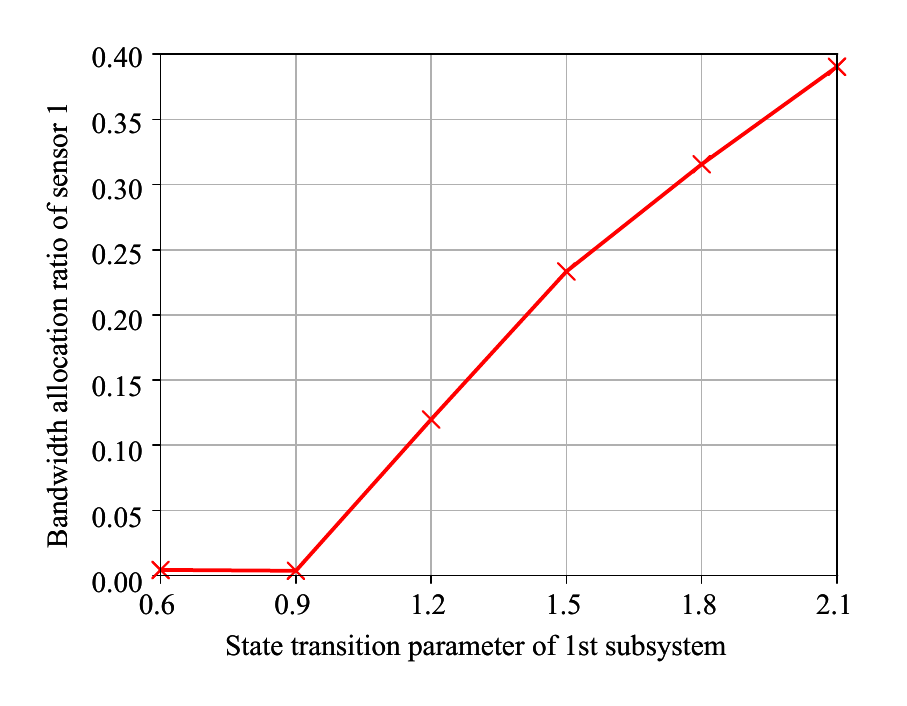}
	\caption{Bandwidth allocation results with different state transition parameters in the decoupled control system scenario.}
	\label{fig:simu_allocation_vs_A}
\end{figure} 

Next, we consider a special scenario characterized by diagonal system matrices $\mathbf{\Phi}$, $\mathbf{\Gamma}$, and $\mathbf{V}$. This structure implies that the elements of system state are decoupled, and the control system can be decomposed into multiple independent subsystems. Specifically, the matrices $\mathbf{\Gamma}$ and $\mathbf{V}$ are set as identical matrices. Fig. \ref{fig:simu_allocation_vs_A} illustrates relationship between the bandwidth allocation ratio for the first subsystem and its corresponding state transition parameter (i.e., the first diagonal element of $\mathbf{\Phi}$), while all other diagonal elements of $\mathbf{\Phi}$ are set to be 1.5. The bandwidth constraint is set as $B_{\text{max}}=50$ kHz. As shown in Fig. \ref{fig:simu_allocation_vs_A}, the bandwidth allocated to the first sensor increases monotonically with its state transition parameter. This behavior is attributed to the fact that a larger state transition parameter corresponds to a less stable open-loop subsystem, thereby requiring more feedback information (and thus more bandwidth) to maintain stability or achieve the desired control performance. This finding aligns with the conclusion in \cite[Proposition 1]{wcl}, which suggests that more communication resources should be allocated to subsystems exhibiting greater instability. In addition, it can be seen that when the state transition parameter is less than 1, indicating an open-loop stable subsystem, negligible bandwidth is allocated to the corresponding sensor. This is because such subsystems are inherently stable and require minimal or no feedback information for stabilization. {Based on the above results, in practical applications, when the system contains multiple relatively independent subsystems, more communication resources should be allocated to the sensors monitoring less stable subsystems, since these subsystems require more accurate feedback information.}

\section{Conclusions}
\label{sec_conclusion}
In this paper, we have investigated a closed-loop control system containing multiple collaborative sensors, a satellite-UAV-enabled EIH, and a robot. In order to improve the control performance of the system, we have formulated an LQR cost minimization problem to optimize the sensor-to-EIH bandwidth allocation. The control inputs have been designed using the Kalman filter and LQR controller. The control process has been reformulated as a POMDP. Accordingly, a novel bandwidth allocation scheme has been proposed based on the DRL method. Simulation results have shown that the proposed DRL-based bandwidth allocation scheme could adapt to varying control parameters, achieving lower LQR cost compared to the traditional communication-oriented schemes.

{The proposed framework also has some limitations. First, the rate-limited distortion in sensing data is approximated as additive Gaussian noise, and the relationship between the transmitted information and the equivalent distortion-noise variance is established based on mutual information. This model is reasonable in the high-dimensional or small-quantization-step cases and can provide a tractable way to characterize the impact of communication rates on sensing quality. However, the quantization distortion of practical sensing data may not strictly follow the independent Gaussian assumption. In addition, the downlink control-command transmission is assumed to be reliable, which is reasonable when the control packets are much shorter than the raw sensing data and the EIH has stronger transmission capability than individual sensors. However, practical downlink transmission may still suffer from packet errors or latency. Moreover, the proposed framework assumes that the bandwidth resource can be allocated as continuous variables, while practical systems often involve discrete transmission bits or quantization levels. Therefore, future work may extend the current framework by considering more practical downlink transmission models and the joint optimization of quantization levels and transmission strategies.}


\newpage

 


\vspace{11pt}


\vfill

\end{document}